# Bidding Machine: Learning to Bid for Directly Optimizing Profits in Display Advertising

Kan Ren, Weinan Zhang, Ke Chang, Yifei Rong, Yong Yu, and Jun Wang

**Abstract**—Real-time bidding (RTB) based display advertising has become one of the key technological advances in computational advertising. RTB enables advertisers to buy individual ad impressions via an auction in real-time and facilitates the evaluation and the bidding of individual impressions across multiple advertisers. In RTB, the advertisers face three main challenges when optimizing their bidding strategies, namely (i) estimating the utility (e.g., conversions, clicks) of the ad impression, (ii) forecasting the market value (thus the cost) of the given ad impression, and (iii) deciding the optimal bid for the given auction based on the first two. Previous solutions assume the first two are solved before addressing the bid optimization problem. However, these challenges are strongly correlated and dealing with any individual problem independently may not be globally optimal. In this paper, we propose *Bidding Machine*, a comprehensive *learning to bid* framework, which consists of three optimizers dealing with each challenge above, and as a whole, jointly optimizes these three parts. We show that such a joint optimization would largely increase the campaign effectiveness and the profit. From the learning perspective, we show that the bidding machine can be updated smoothly with both offline periodical batch or online sequential training schemes. Our extensive offline empirical study and online A/B testing verify the high effectiveness of the proposed bidding machine.

---

## 1 INTRODUCTION

EMERGING in 2009 [28] and popularized since 2011 [12], real-time bidding (RTB) based display advertising has become a major paradigm in computational advertising for both technique and business perspectives. With ad exchanges as intermediaries, RTB enables publishers to sell the individual ad impressions via hosting a real-time auction and facilitates advertisers to evaluate each auctioned ad impression and bid for it.

In RTB display advertising, as is shown is Figure 1, when a user visits one publisher's site e.g., a web page or a mobile app page, (0) a bid request for the corresponding ad display opportunity, along with its information about the underlying user, the domain context and the auction information, (1) is broadcast to hundreds or thousands of advertisers for bid via an ad exchange [39]. With the help of computer algorithms on demand-side platforms (DSPs), each advertiser estimates the potential utility and the possible cost for the received bid request and (2) makes the final decision of the bid price in this real-time auction. Then the ad exchange will (3) determine that the winner, who proposed the highest bid price, could show the ad and pay for the second highest price which is called as the market price [3] (in second-price auction). The whole loop will be finished in less than 100 ms. The winning advertiser (4) would send the ad creative to the user and (5) receive the user response (e.g., click, conversion) later. For each day, such a request-bidding-feedback loop occurs billions of times for an ordinary RTB platform, which makes RTB be a true battlefield of big data. For example, YOYI DSP, which has deployed our proposed algorithms in this paper and hosted the online A/B testing, handles more than 10 billion ad transactions daily in 2017.

In the view of an advertiser, the goal is to spend the campaign budget on the most effective ad opportunities to achieve high profits, which means the ad volume with more positive user responses, e.g., clicks or conversions, yet non-expensive cost. At each time, in order to calculate the bid, the advertiser first predicts the probability of the positive user response of that ad display, i.e., how likely the user is going to click or convert, which is normally measured by the predicted click-through rate (CTR) or conversion rate (CVR). Generally, the advertisrs should bid higher and allocate more budget on the ad inventory with higher CTR or CVR [30]. For most advertisers, the cost, which means the market price of that bid opportunity [3], should be estimated to better determine the bid price.

As we may find in the above description, for each advertiser, there are three main challenges within the bidding procedure. The first is to estimate the *utility*, i.e., CTR or CVR of the ad impression, with the consideration of the user, ad, publisher and contextual information. CTR or CVR qualifies the expected probability of click or conversion for the advertiser w.r.t. the given ad request. The second challenge is to forecast the probably *cost* for showing the ad, which is the amount paid for the impression. Note that the true cost is the market price rather than the bid price for the winner of the auction. The last but the most important problem is to adopt a proper *bidding strategy* to win as many effective (high utility, yet low cost) impressions as possible to maximize the profits of the advertiser with the campaign budget constraint.

- K. Ren, W. Zhang, K. Chang and Y. Yu are with Shanghai Jiao Tong University. E-mail: {kren, kchang, wnzhang, yyu}@apex.sjtu.edu.cn
- Y. Rong is with Meituan-Dianping Inc. E-mail: rongyifei@meituan.com. This work was done when he was with YOYI Inc.
- J. Wang is with University College London.
  E-mail: jun.wang@cs.ucl.ac.uk

Typically, the bid optimization is done in the following sequential basis. First, the CTR estimation[1] is formulated as a binary regression problem, which can be solved by machine learning models such as logistic regressions [25], [35], Bayesian probit regression [13], gradient boosting regression trees [14], factorization machines [29], etc. The common objective in this stage is to make the estimation as accurate as possible by, for instance, minimizing the cross entropy error between the predicted CTR and ground truth user responses. Second, for cost estimation, literatures [9], [41] formulate the problem as a prediction task to forecast the market price distribution, which is named as bid landscape forecasting, or directly estimate the winning price of the given bid request. Third, based on the estimated CTR and cost of the ad display opportunity, we will seek for the optimal bidding function along with other considerations including the campaign budget and the auction volume etc. [19], [22], [30], [47], [48].

However, such sequential optimization is indeed not optimal. According to the Bayesian decision theory [4], the learning of the user response model and bid landscape model should be informed by the final bidding utility. As is studied in the literature [33], the required accuracy of the CTR prediction would not be the same throughout the range of the prediction $[0, 1]$ as there is a cost (negative utility) for the advertiser to win an impression if no click, but no cost (zero utility) for losing one. The value of clicks also varies across campaigns; and it would be good if the CTR learning can tailor its efforts more toward those higher-valued cases and make them better predicted. More importantly, the user response prediction is indeed correlated with the second price auction in RTB — if won an auction, the advertiser pays the market price and then obtain the payoff from the user conversions triggered by the ad.

Therefore, the market price and the competition have a significant impact on the campaign performance. On one hand, if the performed bid is in a highly competitive situation, it is of low confidence to predict whether the advertiser will win the ad auction or not; thus the optimization of the CTR prediction in such case should be more focused and fine-tuned than that in the less competitive case. On the other hand, the bidding strategy also influences the utility estimation and cost prediction modules. In [22], the authors proved the advantage for bidding effectiveness of combining the two optimization models for both CTR and the winning price under budget constraints. As a consequence, the natural idea is to solve the three main challenges altogether and derive a comprehensive methodology to jointly optimize the bidding performance.

In this paper, we present a novel optimization framework, named as Bidding Machine (BM) as shown in Figure 1, which considers the three challenges as a whole and directly pushes the limit of the campaign profit by jointly optimizing the three components: user response prediction, bid landscape forecasting and bid optimization. In bidding machine, we adopt the utility estimation model and the cost prediction model while utilizing a comprehensive utility optimization objective function. Moreover, we take the budget into consideration and prove the optimal bidding strategy under the second price auction. Whereafter a functional optimization method is used to derive the optimal bidding function with budget constraints. The overall methodology works as a learning to bid model, which consumes recent historical bidding logs and (i) updates the estimation models and (ii) optimizes the corresponding bidding strategy; then (iii) performs the bidding phase online and observes the real-world feedback. After that, it returns to the first step and repeats the *update-optimize-perform* loop. The procedure circulates and acts as a machine interacting with the market while trying to maximize the obtained profits.

---
1. In this paper, we focus on the CTR estimation, while the CVR estimation can be done by following the same token.

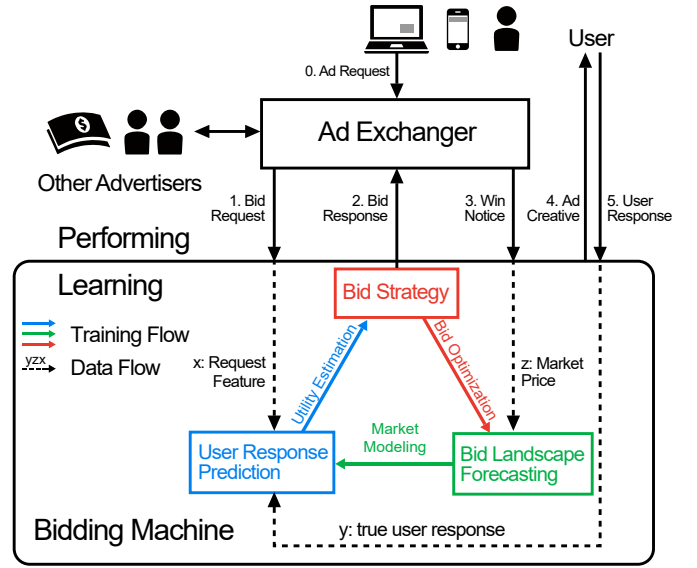

Fig. 1: The joint learning framework of bidding machine.

Note that our methodology tries to resolve the three main challenges in a comprehensive optimization framework and directly optimize the profit for the advertiser. To our best knowledge, it is the first work that takes these three key components of RTB altogether to optimize. In our recent work [33], we combined utility estimation and bidding strategy in a whole while adopting naive counting-based method for market price modeling, which is not optimal considering the expected cost since the market price is flexible in different contexts. The authors in [22] proposed a method which combines CTR estimation and winning price prediction (as will be compared in the experiment). But the method still regards the two aspects as separate problems and does not put them into a joint optimization framework.

To sum up, the contributions of this paper are three-fold: (i) We point out the three main challenges in RTB, namely user response prediction, bid landscape forecasting and bid optimization, are indeed highly correlated but are commonly tackled separately in previous work. (ii) We propose bidding machine, a comprehensive framework to jointly optimize these three components to directly push the limit of the campaign profit. (iii) Extensive empirical study demonstrates that BM vastly improves the bidding performance against the state-of-the-art baselines, in both offline experiments on two public datasets and online A/B testing on a commercial RTB platform.

In the rest of our paper, we first discuss related literatures in Sec. 2 and then present the bidding machine framework. Specifically, we formulate the problem in Sec. 3, and then present our CTR estimation model in Sec. 4 and cost prediction model in Sec. 5. After that, we solve the functional optimization problem for the optimal bidding strategy in Sec. 6. Due to page limit, the online learning version is presented in Appendix A, also with the proof of the optimal bidding function under the second-price auction described in Appendix B. We also make a game theoretic analysis of multiple bidders with the same optimal strategy in Appendix C. We present the experiments and discuss the detailed results in Sec. 7. Finally we conclude this paper and discuss the future work in Sec. 8.

## 2 RELATED WORK

In this section, we discuss the related literatures about RTB techniques, specifically the three key components of RTB, as pointed out in Sec. 1, namely utility estimation, cost prediction and bidding strategy optimization.

**Utility: User Response Prediction.** The first challenge is the utility estimation, which is mostly about user response prediction, such as the click-through rate (CTR) estimation or the conversion rate (CVR) estimation. And it plays a key role in real-time display advertising [19], [25], [38]. The response prediction is a probability estimation task [27] which models the interest of users in the content of publishers or the ads, and is used to derive the budget allocation of the advertisers [35]. Typically, the response prediction problem is formulated as a binary regression problem with prediction likelihood as the training objective [1], [13], [29], [35]. From the methodology view, linear models such as logistic regression [19] and non-linear models such as tree-based model [14] and factorization machines [27], [29] are commonly used. Other variants include Bayesian probit regression [13], FTRL-trained factorization machine [36], and neural network learning framework [31]. Normally, area under ROC curve (AUC) and relative information gain (RIG) are common evaluation metrics for CTR prediction accuracy [13]. Recently, the authors in [7], [37] pointed out that such metrics may not be good enough for evaluating CTR predictor in RTB based advertising because of the subsequent bidding and auctions. The authors in [37] proposed a cost-sensitive objective function to tackle the cost issue in user response prediction learning. However, it may be influenced by the suboptimal cost estimation module. In this part of the paper, considering utility estimation, we use a logistic regression as a working example and go one step further over [7] to reformulate the CTR estimation learning by directly optimizing campaign performance (profit).

**Cost: Bid Landscape Forecasting.** For cost estimation, we refer it to bid landscape forecasting, which aims at predicting the distribution of market price for a type of ad inventory [9]. The advertisers use it to calculate the winning rate given a bid and help decide the final bid price. Several winning function forms were hypothesized in [20], [48] to directly induce the optimal bidding functions. A campaign-level forecasting system with tree models was presented in [9]. The authors in [17] conducted an error handling methodology to improve the efficiency and reliability of the bid landscape forecasting system. As advertisers only know the statistics (market price, user clicks etc.) from their winning impressions, the authors in [41] proposed a solution to handle such data censorship in market price prediction. Later we will show that market price distribution indeed plays an essential role in both CTR model learning and bidding strategy optimization for campaign profit optimization, which has never been formally discussed.

**Strategy: Bid Optimization.** With the estimated utility of CTR/CVR, the advertisers would be able to assess the value of the impression and perform a bid. The auction theory [11] proves that truthful bidding, i.e., bidding the action value times the action rate, is the optimal strategy in the second price auction [19]. However, with budget and auction volume constraints, the truthful bidding may not be optimal [46]. The linear bidding strategy [30] is widely used in industry, where the bid price is calculated via the predicted CTR/CVR multiplied by a constant parameter tuned according to the campaign budget and performance. The authors in [8] proposed a bidding function with truthful bidding value minus a tuned parameter. A lift-based bidding strategy was recently proposed in [43] where the bid price was determined by the user CVR lift after seeing the displayed ad.

However, the impact of market price distribution, i.e., bid landscape, has not yet been studied in the above works, and the final utility of the campaign is not considered in the optimization objective, which may result in some unfavorable statistics such as relatively high effective cost per click (eCPC) and low return-on-investment ratio (ROI). The authors in [20] combined the winning rate estimation and the winning price prediction together and deployed the estimation results in different bidding strategies towards different business demands. The authors in [18] embedded a budget smoothing component into a bid optimization framework. In [47], [48], with the estimated CTR as input of the bidding function, the authors leveraged functional optimization to derive non-linear bidding functions. Our work is different from the above works as we directly model CTR learning and cost estimation as part of bid optimization for campaign profit maximization. In [22] the authors combined two predictors to decide the final bid price. However, the proposed model may be suboptimal since they did not take profit as the objective function and heuristically set the bidding as a predicted winning price plus a constant value.

To sum up, all the existing learning frameworks in RTB consider the user response prediction, bid landscape forecasting and bid optimization as three separated parts, while in our paper, we model them as a whole and perform a novel joint optimization.

## 3 PROBLEM DEFINITION OF BIDDING MACHINE

We propose a unified learning framework, named as *Bidding Machine*, which integrates both utility and cost estimation and puts them back into bid optimization, to maximize the overall profit for advertisers.

Recall that in RTB scenario, on one side, the user visiting an online page may trigger an auction, in real time, for each ad slot on the web page. On the other side, the advertiser receives a bid request from the ad exchange [45], along with the information about user, context on web page and other auction features. Then the bidding engine of DSP decides, on behalf of a campaign, whether to participate the auction and how much to bid on this impression opportunity.

As is discussed in many literatures [8], [22], [30], [48], the final bid price is influenced by many factors including estimated utility (i.e., user response) and cost (i.e., market price). And bid optimization should also consider the budget pacing [2] to control the cost to optimize the final profit. However, almost all the related work bases bidding decision on the estimated utility and the predicted cost, and treats these three parts desperately. [33] embedded user response prediction, i.e., utility estimation, into the bid optimization and made the prediction aware of the market information and cost sensitivity. [48] took budget constraint, winning probability and cost of the particular ad impression altogether and proposed a functional bidding optimization to maximize the target KPIs, e.g., clicks.

Our work goes steps further. We propose a unified optimization framework on the integral utility minus cost objective, which is the profit gained by the advertiser. Our goal is to learn the user response prediction model with market competition modeling, and optimize the final bidding strategy considering the preset budget constraints. We will first formulate the unified learning problem and then discuss our optimization solutions. Note that the derived learning formulations of each component benefit from the update of the other two parts. Thus the whole optimization framework is a joint learning procedure, where the user response learning, market competition modeling and bidding strategy optimization run as a circulation mechanism.

## 3.1 Problem Definition

Typically, a bid request contains various information of an ad display opportunity, including the information of the underlying user, location, time, browser and other contextual information about the web page. Along with the features extracted from the campaign itself, we construct the high-dimensional feature vector for the bid request, which is denoted as $\boldsymbol{x}$. We also use $p_{\boldsymbol{x}}(\boldsymbol{x})$ to denote the probability distribution of the input feature vector $\boldsymbol{x}$ that matches the campaign target rules.

Without loss of generality, we take click as user response and the CTR estimation is denoted as a function $p(y = 1|\boldsymbol{x}) \equiv f_{\boldsymbol{\theta}}(\boldsymbol{x})$ mapping from feature $\boldsymbol{x}$ to the probability of a click, where $y \in \{0, 1\}$ is a binary variable indicating whether a user click occurs (1) or not (0). We define the true value of an occurring click as $v$ which is preset by the advertiser.

Next, we define the context where utility estimation is situated. A lot of previous work has specified the bidding strategy as a function $b(f_{\boldsymbol{\theta}}(\boldsymbol{x}))$ mapping from the predicted CTR (or other estimated KPIs) $f_{\boldsymbol{\theta}}(\boldsymbol{x})$ to the bid price [19], [30], [48]. Essentially, the mapping follows a sequential dependency assumption $\boldsymbol{x} \to f_{\boldsymbol{\theta}}(\boldsymbol{x}) \to b$ proposed by [47], [48]. In this paper, we follow the same formulation. For simplicity, we use $b(\cdot)$ to represent the bidding function, but also occasionally use $b$ to directly represent the bid price.

Once the DSP sends out the bid $b$, the ad exchange hosts a second-price auction [15] and decides who is going to win the auction. The probability of winning an auction is influenced by the bid price $b$ and the stochastic market price $z$ with an underlying p.d.f. $p_z(z)$; we use $w_{\boldsymbol{\phi}}(b)$ to denote the probability of winning as:

$$w_{\boldsymbol{\phi}}(b) = \int_0^b p_z(z)dz, \quad (1)$$

which is the probability that the bid $b$ is higher than the market price $z$ [15] and $\boldsymbol{\phi}$ is the parameter of our winning probablity model. The details of the winning function will be discussed later.

TABLE 1: Notations and descriptions

| Notation | Description |
|---|---|
| $v$ | The pre-defined value of positive user response. |
| $y$ | The true label of user response. |
| $\boldsymbol{x}$ | The bid request represented by its features. |
| $p_{\boldsymbol{x}}(\boldsymbol{x})$ | The probabilistic density function of $\boldsymbol{x}$. |
| $z$ | The market price. |
| $p_z(z)$ | The probabilistic density function of $z$. |
| $\boldsymbol{\theta}$ | The weight of CTR estimation function. |
| $f_{\boldsymbol{\theta}}(\boldsymbol{x})$ | The CTR estimation function to learn. |
| $r$ | The predicted CTR. |
| $b(f_{\boldsymbol{\theta}}(\boldsymbol{x}))$ | The bid price determined by the estimated CTR, $b$ for short. |
| $D$ | The training set. |
| $R(\cdot)$ | The utility function. |
| $w_{\boldsymbol{\phi}}(b)$ | The winning probability given bid price $b$. |
| $c(b)$ | The expected cost given bid price $b$ if winning. |

If the bid wins the auction, the advertiser pays the cost, which is the market price $z$. We denote the expected cost in the second price auction as

$$c(b) = \frac{\int_0^b z p_z(z)dz}{\int_0^b p_z(z)dz}, \quad (2)$$

which is essentially the expected market price when winning the auction [15]. Once we have defined the bidding function $b$, the true value of a click $v$, and the winning rate $w$, the expected cost $c$, we are ready to define a general form of the utility function as $R_{\boldsymbol{\theta}}(\boldsymbol{x}, y; b, v, c, w)$ for a given $(\boldsymbol{x}, y)$ 2-tuple in the training data (all the received historical impressions).

Our task is to build a joint optimization framework modeling user response, market competition and bidding strategy, to maximize the overall profit for the advertiser, which is formulated as

$$(b(\cdot), \boldsymbol{\theta}^*, \boldsymbol{\phi}^*) = \arg\max_{b, \boldsymbol{\theta}, \boldsymbol{\phi}} \int_{\boldsymbol{x}} R(\boldsymbol{x}, y; b, v, c, w) p_{\boldsymbol{x}}(\boldsymbol{x}) d\boldsymbol{x}, \quad (3)$$

Following [7], [33], the overall revenue $R(\cdot)$ can be defined as *utility function* below w.r.t. the corresponding auction sample $(\boldsymbol{x}, y)$:

$$\begin{aligned} R(b, \boldsymbol{\theta}, \boldsymbol{\phi}) &= \int_{\boldsymbol{x}} [vy - c(b(f_{\boldsymbol{\theta}}(\boldsymbol{x})))] w_{\boldsymbol{\phi}}(b(f_{\boldsymbol{\theta}}(\boldsymbol{x}))) p_{\boldsymbol{x}}(\boldsymbol{x}) d\boldsymbol{x} \\ &= \sum_{(\boldsymbol{x}, y) \in D} [vy - c(b(f_{\boldsymbol{\theta}}(\boldsymbol{x})))] w_{\boldsymbol{\phi}}(b(f_{\boldsymbol{\theta}}(\boldsymbol{x}))). \end{aligned} \quad (4)$$

The summary of our notations are gathered in Table 1. In the next three sections, we will discuss the detailed optimization of $R(b, \boldsymbol{\theta}, \boldsymbol{\phi})$ w.r.t. the CTR estimator $f_{\boldsymbol{\theta}}(\cdot)$ (Sec. 4), the winning function $w_{\boldsymbol{\phi}}(\cdot)$ (Sec. 5) and the bidding function $b(\cdot)$ (Sec. 6).

## 4 UTILITY: USER RESPONSE LEARNING

In this section, we formally describe the utility estimation task, i.e., user response learning, and propose two solutions for this problem. Recall that, in the training set defined as $D$, each sample is represented as a 2-tuple as $(\boldsymbol{x}, y)$, where $\boldsymbol{x}$ denotes the feature vector of the bid request, and $y$ denotes the indicator whether user action (click) occurs.

### 4.1 Gradient for Expected Utility

To solve Eq. (3), utility function $R_{\boldsymbol{\theta}}(\cdot)$ can be naturally defined as the expected direct profit from the campaign:

$$R^{\text{EU}}_{\boldsymbol{\theta}}(\boldsymbol{x}, y) = [vy - c(b(f_{\boldsymbol{\theta}}(\boldsymbol{x})))] \cdot w(b(f_{\boldsymbol{\theta}}(\boldsymbol{x}))), \quad (5)$$

where to simplify our notation, we drop the dependency of $b, v, c, w$ for $R^{\text{EU}}_{\boldsymbol{\theta}}(\boldsymbol{x}, y)$. The expectation is w.r.t. whether winning or not, where no winning has zero utility. Recall that, in the training set defined as $D$, each sample is represented as a 2-tuple as $(\boldsymbol{x}, y)$, where $\boldsymbol{x}$ denotes the feature vector of the bid request, and $y$ denotes the indicator whether user action (click) occurs. The overall expected direct profit [7] of all the auctions can be calculated by replacing Eqs. (1) and (2) into Eq. (5) as

$$\begin{aligned}
&\sum_{(\boldsymbol{x},y) \in D} R^{\text{EU}}_{\boldsymbol{\theta}}(\boldsymbol{x}, y) \\
&= \sum_{(\boldsymbol{x},y) \in D} \left[ vy - \frac{\int_0^{b(f_{\boldsymbol{\theta}}(\boldsymbol{x}))} z \cdot p_z(z) dz}{\int_0^{b(f_{\boldsymbol{\theta}}(\boldsymbol{x}))} p_z(z) dz} \right] \cdot \int_0^{b(f_{\boldsymbol{\theta}}(\boldsymbol{x}))} p_z(z) dz \\
&= \sum_{(\boldsymbol{x},y) \in D} \int_0^{b(f_{\boldsymbol{\theta}}(\boldsymbol{x}))} (vy - z) \cdot p_z(z) dz. \quad (6)
\end{aligned}$$

Taking Eq. (6) into Eq. (3) with a regularization term, we have

$$\begin{aligned}
\boldsymbol{\theta}^{\text{EU}} &= \arg\min_{\boldsymbol{\theta}} \; - \sum_{(\boldsymbol{x},y) \in D} R^{\text{EU}}_{\boldsymbol{\theta}}(\boldsymbol{x}, y) + \frac{\lambda}{2} \|\boldsymbol{\theta}\|_2^2 \quad (7) \\
&= \arg\min_{\boldsymbol{\theta}} \; - \sum_{\boldsymbol{x}} \int_0^{b(f_{\boldsymbol{\theta}}(\boldsymbol{x}))} (vy - z) \cdot p_z(z) dz + \frac{\lambda}{2} \boldsymbol{\theta}^T \boldsymbol{\theta},
\end{aligned}$$

where the optimal value of $\boldsymbol{\theta}$ is obtained by taking a gradient descent algorithm. The gradient of $R^{\text{EU}}_{\boldsymbol{\theta}}(\boldsymbol{x}, y)$ with regard to $\boldsymbol{\theta}$ is calculated as

$$\frac{\partial R^{\text{EU}}_{\boldsymbol{\theta}}(\boldsymbol{x}, y)}{\partial \boldsymbol{\theta}} = \overbrace{(vy - b(f_{\boldsymbol{\theta}}(\boldsymbol{x})))}^{\text{bid error}} \cdot \overbrace{p_z(b(f_{\boldsymbol{\theta}}(\boldsymbol{x})))}^{\text{market sensitivity}} \cdot \frac{\partial b(f_{\boldsymbol{\theta}}(\boldsymbol{x}))}{\partial f_{\boldsymbol{\theta}}(\boldsymbol{x})} \frac{\partial f_{\boldsymbol{\theta}}(\boldsymbol{x})}{\partial \boldsymbol{\theta}}, \quad (8)$$

and we update for each data instance as $\boldsymbol{\theta} \leftarrow \boldsymbol{\theta} - \eta \left( -\frac{\partial R^{\text{EU}}_{\boldsymbol{\theta}}(\boldsymbol{x},y)}{\partial \boldsymbol{\theta}} + \lambda \boldsymbol{\theta} \right)$ with above chain rule where $\eta$ is the learning rate.

We take logistic regression as our CTR prediction function, i.e.,

$$f_{\boldsymbol{\theta}}(\boldsymbol{x}) \equiv \sigma(\boldsymbol{\theta}^T \boldsymbol{x}) = \frac{1}{1 + e^{-\boldsymbol{\theta}^T \boldsymbol{x}}}, \quad (9)$$

and get $\frac{\partial f_{\boldsymbol{\theta}}(\boldsymbol{x})}{\partial \boldsymbol{\theta}} = \sigma(\boldsymbol{\theta}^T \boldsymbol{x})(1 - \sigma(\boldsymbol{\theta}^T \boldsymbol{x}))\boldsymbol{x}$. After learning, we denote the predicted CTR as $r = f_{\boldsymbol{\theta}}(\boldsymbol{x})$ to simplify the subsequent derivation. We will discuss the specific formulation and the corresponding optimization method for bidding function $b(r)$ w.r.t. the predicted CTR $r$.

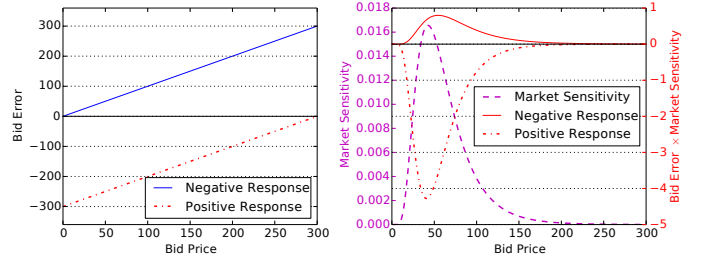

Fig. 2: The illustration of the impact from the bid and market price of Expected Utility (EU); click value $v = 300$.

**Discussion.** Eq. (8) provides a novel gradient update, taking into account both the utility and the cost of a bidding decision (the *bid error* term) as well as the impact from the market price distribution (the *market sensitivity* term). They act as two additional re-weighting functions influencing a conventional gradient update, which is formulated by the remaining terms in the equation. We illustrate their impact in Figure 2. The left subfigure shows the weight from *bid error* against bid price with different user responses ($y = 1$ or $y = 0$). We see that the update of the CTR model aims to correct a bid towards the true value $vy$ from a training instance, i.e., an optimal model (parameter) would generate a bid close to $v$ for a positive instance, while close to zero for a negative instance. The right subfigure plots the weight adjustment from the *market sensitivity* term (y-axis left) and the combined weight *bid error* × *market sensitivity* (y-axis right). We observe that the *market sensitivity* term re-weights the bid error by checking the fitness to the market price distribution; this makes the gradient focused more on fixing the errors (if any) when the bid is close to the market price. This is intuitively correct because when the bid is close to the market price, the competition is high and a small error (win a case that is no click and vice versa) would make a huge difference in terms of the cost and reward. Specifically, for the negative response ($y = 0$), the combined weight $bp_z(b)$ stays positive in order to constantly lower the bid via CTR learning, but its peak location is slightly higher than the mode of market price. For the positive response ($y = 1$), the combining weight $(b - v)p_z(b)$ in $-\frac{\partial R^{\text{EU}}_{\boldsymbol{\theta}}(\boldsymbol{x},y)}{\partial \boldsymbol{\theta}}$ is negative to push the bid higher to $v$. Note that the bid is restricted in $[0, v]$ as bidding higher than $v$ is of no advantage than bidding $v$ when optimizing profit.

### 4.2 Gradient for Risk-Return

Besides the expected utility (EU), we also propose a risk-return (RR) model to balance the risk and return of a bid decision as below:

$$R^{\text{RR}}_{\boldsymbol{\theta}}(\boldsymbol{x}, y) = \Big( \underbrace{\frac{vy}{z}}_{\text{return}} - \underbrace{\frac{v(1-y)}{v-z}}_{\text{risk}} \Big) \cdot w(b(f_{\boldsymbol{\theta}}(\boldsymbol{x}))), \quad (10)$$

where we define that when $y = 1$, the winning utility is $\frac{v}{z}$, which is the ratio between the return and the cost of this transaction; when $y = 0$, the winning utility becomes the penalty for taking risk $\frac{-v}{v-z}$, which is defined as the ratio between the lost ($-v$) and the gain if winning ($v - z$). Note that $v$ is always higher than $z$ as $v \geq b > z$. The penalty is very high when bidding for a very low margin (low $v - z$) case. Thus the new optimization objective function is

$$\begin{aligned}\boldsymbol{\theta}^{\text{RR}} &= \arg\min_{\boldsymbol{\theta}} \ -\sum_{(\boldsymbol{x},y)\in D} R_{\boldsymbol{\theta}}^{\text{RR}}(\boldsymbol{x},y) + \frac{\lambda}{2}\|\boldsymbol{\theta}\|_2^2 \\ &= \arg\min_{\boldsymbol{\theta}} \ -\sum_{(\boldsymbol{x},y)\in D}\int_0^{b(f_{\boldsymbol{\theta}}(\boldsymbol{x}))}\Big(\frac{vy}{z} - \frac{v(1-y)}{v-z}\Big) p_z(z) dz \\ &\quad + \frac{\lambda}{2}\boldsymbol{\theta}^T\boldsymbol{\theta},\end{aligned} \quad (11)$$

which leads to the gradient of $R_{\boldsymbol{\theta}}^{\text{RR}}(\boldsymbol{x}, y)$ w.r.t. $\boldsymbol{\theta}$ as

$$\frac{\partial R_{\boldsymbol{\theta}}^{\text{RR}}(\boldsymbol{x},y)}{\partial \boldsymbol{\theta}} = \Big(\overbrace{\frac{vy}{b(f_{\boldsymbol{\theta}}(\boldsymbol{x}))} - \frac{v(1-y)}{v - b(f_{\boldsymbol{\theta}}(\boldsymbol{x}))}}^{\text{bid error}}\Big) \cdot \overbrace{p_z(b(f_{\boldsymbol{\theta}}(\boldsymbol{x})))}^{\text{market sensitivity}} \\ \cdot \frac{\partial b(f_{\boldsymbol{\theta}}(\boldsymbol{x}))}{\partial f_{\boldsymbol{\theta}}(\boldsymbol{x})} \frac{\partial f_{\boldsymbol{\theta}}(\boldsymbol{x})}{\partial \boldsymbol{\theta}} \ . \quad (12)$$

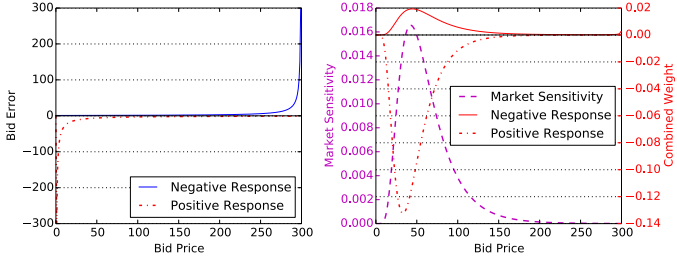

Fig. 3: The illustration of the impact from the bid and market price of Risk Return (RR); click value $v = 300$.

**Discussion.** To understand the above gradient, we plot the *bid error*, *market sensitivity* and their combined weight in Figure 3. The RR model is different from the previous EU model in that, the *bid error* turns to *return* when the response is positive, and becomes *risk* when meets a negative response. If $y = 0$ and bid price is high, or if $y = 1$ and bid price is low, the bid error is quite significant to avoid the happening of such cases.

As is shown in both Eqs. (8) and (12), the market price distribution plays an important role in the optimization: with the determined bidding function and CTR estimation function, the gradient is weighted by the probabilistic density function (p.d.f.) of market price, denoted as $p_z(z)$.

### 4.3 Model Realization

Solving the proposed learning objectives (7) and (11) relies on the realization of the bidding function $b(f_{\boldsymbol{\theta}}(\boldsymbol{x}))$, the market price distribution $p_z(z)$ and the CTR estimation function itself $f_{\boldsymbol{\theta}}(\boldsymbol{x})$. In this section, we will discuss the solutions from the proposed two training objectives given some specific implementations of $b(f_{\boldsymbol{\theta}}(\boldsymbol{x}))$, $p_z(z)$ and $f_{\boldsymbol{\theta}}(\boldsymbol{x})$.

Without loss of generality, for the CTR estimation model, we adopt the widely used logistic regression for $f_{\boldsymbol{\theta}}(\boldsymbol{x})$ as in Eq. (9).

For the bidding strategy, we employ a widely used linear bidding function w.r.t. the predicted CTR [30] with a scaling parameter $\rho$

$$b(f_{\boldsymbol{\theta}}(\boldsymbol{x})) \equiv \rho \cdot v \cdot f_{\boldsymbol{\theta}}(\boldsymbol{x}). \quad (13)$$

Taking Eqs. (9) and (13) into (8) and (12), respectively, we derive our final gradient of the proposed EU utility:

$$\frac{\partial R_{\boldsymbol{\theta}}^{\text{EU}}(\boldsymbol{x},y)}{\partial \boldsymbol{\theta}} = \rho v^2(y - \rho\sigma(\boldsymbol{\theta}^T\boldsymbol{x})) \cdot p_z(b(f_{\boldsymbol{\theta}}(\boldsymbol{x}))) \cdot \\ \sigma(\boldsymbol{\theta}^T\boldsymbol{x})(1 - \sigma(\boldsymbol{\theta}^T\boldsymbol{x}))\boldsymbol{x} \ , \quad (14)$$

and that of the RR utility:

$$\frac{\partial R_{\boldsymbol{\theta}}^{\text{RR}}(\boldsymbol{x},y)}{\partial \boldsymbol{\theta}} = \rho v\Big(\frac{y}{\rho\sigma(\boldsymbol{\theta}^T\boldsymbol{x})} - \frac{1-y}{1 - \rho\sigma(\boldsymbol{\theta}^T\boldsymbol{x})}\Big) \cdot \\ p_z(b(f_{\boldsymbol{\theta}}(\boldsymbol{x}))) \cdot \sigma(\boldsymbol{\theta}^T\boldsymbol{x})(1 - \sigma(\boldsymbol{\theta}^T\boldsymbol{x}))\boldsymbol{x} \ , \quad (15)$$

where the bidding function parameter $\rho$ acts as a calibration term in bid correction.

Note that, various bid landscape models can be utilized to model $p_z(z)$, such as the parametric log-normal distribution [9] and Gamma distribution [7]. In this paper, while our model is flexible with various landscape models, we first adopt a non-parametric $p_z(z)$ which is directly obtained from each campaign's winning price data [3], and also a parametric functional model $p_z(z, \boldsymbol{x}; \boldsymbol{\phi})$ which will be discussed later.

### 4.4 Links to Previous Work

It is of great interest to compare our profit-optimized solutions with the existing ones that optimize the fitness of the user response data. A logistic regression could be trained with squared error (SE) loss to fit user response data:

$$\mathcal{L}_{\boldsymbol{\theta}}^{\text{SE}}(\boldsymbol{x},y) = \frac{1}{2}(y - \sigma(\boldsymbol{\theta}^T\boldsymbol{x}))^2, \\ \frac{\partial \mathcal{L}_{\boldsymbol{\theta}}^{\text{SE}}(\boldsymbol{x},y)}{\partial \boldsymbol{\theta}} = (\sigma(\boldsymbol{\theta}^T\boldsymbol{x}) - y)\sigma(\boldsymbol{\theta}^T\boldsymbol{x})(1 - \sigma(\boldsymbol{\theta}^T\boldsymbol{x}))\boldsymbol{x}. \quad (16)$$

More commonly, in a binary output case, a logistic regression can be also trained with cross entropy (CE) loss:

$$\mathcal{L}_{\boldsymbol{\theta}}^{\text{CE}}(\boldsymbol{x},y) = -y\log\sigma(\boldsymbol{\theta}^T\boldsymbol{x}) - (1-y)\log(1 - \sigma(\boldsymbol{\theta}^T\boldsymbol{x})), \\ \frac{\partial \mathcal{L}_{\boldsymbol{\theta}}^{\text{CE}}(\boldsymbol{x},y)}{\partial \boldsymbol{\theta}} = (\sigma(\boldsymbol{\theta}^T\boldsymbol{x}) - y)\boldsymbol{x}. \quad (17)$$

We see that our solutions in Eq. (14) and Eq. (15) extend the original gradients in Eq. (16) and Eq. (17) by (i) replacing the user response errors with the bid errors, and (ii) adding the consideration from the market price and the competition. Under the assumption of (i) truthful bidding function [19], [30] and (ii) uniform market price distribution as

$$b(f_{\boldsymbol{\theta}}(\boldsymbol{x})) = v \cdot f_{\boldsymbol{\theta}}(\boldsymbol{x}) \ , \quad (18)$$
$$p_z(z) = l \ , \quad (19)$$

our proposed learning models which directly optimize (maximize) the profit-related utility are equivalent to the standard logistic regression with (minimizing) squared error loss or cross entropy loss:

$$\frac{-\partial R_{\boldsymbol{\theta}}^{\text{EU}}(\boldsymbol{x},y)}{\partial \boldsymbol{\theta}} = v^2 l(\sigma(\boldsymbol{\theta}^T\boldsymbol{x}) - y) \cdot \sigma(\boldsymbol{\theta}^T\boldsymbol{x})(1 - \sigma(\boldsymbol{\theta}^T\boldsymbol{x}))\boldsymbol{x} \ , \\ \frac{-\partial R_{\boldsymbol{\theta}}^{\text{RR}}(\boldsymbol{x},y)}{\partial \boldsymbol{\theta}} = vl(\sigma(\boldsymbol{\theta}^T\boldsymbol{x}) - y)\boldsymbol{x} \ . \quad (20)$$

Table 2 summarizes and provides a straightforward comparison among various model settings with the EU and RR loss. But note that, in our settings, we adopt more reasonable bidding functions and market price distributions, to achieve substantial improvement against the traditional regression loss methods.

TABLE 2: The comparison of the model gradients (without regularization). LR: logistic regression, TB: truthful bidding, LB: linear bidding, UM: uniform market price distribution. LR and LR+TB+UM are equivalent (LR+TB reduces to the baseline LR when assuming the uniform market price distribution).

| Model Setting | EU (SE) Gradient | RR (CE) Gradient |
| --- | --- | --- |
| LR (baseline) | $\frac{\partial \mathcal{L}^{\text{SE}}_{\boldsymbol{\theta}}(\boldsymbol{x},y)}{\partial \boldsymbol{\theta}} = (\sigma(\boldsymbol{\theta}^T\boldsymbol{x}) - y) \cdot \sigma(\boldsymbol{\theta}^T\boldsymbol{x})(1 - \sigma(\boldsymbol{\theta}^T\boldsymbol{x}))\boldsymbol{x}$ | $\frac{\partial \mathcal{L}^{\text{CE}}_{\boldsymbol{\theta}}(\boldsymbol{x},y)}{\partial \boldsymbol{\theta}} = (\sigma(\boldsymbol{\theta}^T\boldsymbol{x}) - y)\boldsymbol{x}$ |
| LR+TB | $-\frac{\partial R^{\text{EU}}_{\boldsymbol{\theta}}(\boldsymbol{x},y)}{\partial \boldsymbol{\theta}} = v^2(\sigma(\boldsymbol{\theta}^T\boldsymbol{x}) - y) \cdot p_z(b(f_{\boldsymbol{\theta}}(\boldsymbol{x}))) \cdot \sigma(\boldsymbol{\theta}^T\boldsymbol{x})(1 - \sigma(\boldsymbol{\theta}^T\boldsymbol{x}))\boldsymbol{x}$ | $-\frac{\partial R^{\text{RR}}_{\boldsymbol{\theta}}(\boldsymbol{x},y)}{\partial \boldsymbol{\theta}} = v(\sigma(\boldsymbol{\theta}^T\boldsymbol{x}) - y) \cdot p_z(b(f_{\boldsymbol{\theta}}(\boldsymbol{x}))) \cdot \boldsymbol{x}$ |
| LR+TB+UM | $-\frac{\partial R^{\text{EU}}_{\boldsymbol{\theta}}(\boldsymbol{x},y)}{\partial \boldsymbol{\theta}} = v^2 l(\sigma(\boldsymbol{\theta}^T\boldsymbol{x}) - y) \cdot \sigma(\boldsymbol{\theta}^T\boldsymbol{x})(1 - \sigma(\boldsymbol{\theta}^T\boldsymbol{x}))\boldsymbol{x}$ | $-\frac{\partial R^{\text{RR}}_{\boldsymbol{\theta}}(\boldsymbol{x},y)}{\partial \boldsymbol{\theta}} = v l(\sigma(\boldsymbol{\theta}^T\boldsymbol{x}) - y)\boldsymbol{x}$ |
| LR+LB | $-\frac{\partial R^{\text{EU}}_{\boldsymbol{\theta}}(\boldsymbol{x},y)}{\partial \boldsymbol{\theta}} = \phi v^2 (\phi\sigma(\boldsymbol{\theta}^T\boldsymbol{x}) - y) \cdot p_z(b(f_{\boldsymbol{\theta}}(\boldsymbol{x}))) \cdot \sigma(\boldsymbol{\theta}^T\boldsymbol{x})(1 - \sigma(\boldsymbol{\theta}^T\boldsymbol{x}))\boldsymbol{x}$ | $-\frac{\partial R^{\text{RR}}_{\boldsymbol{\theta}}(\boldsymbol{x},y)}{\partial \boldsymbol{\theta}} = \phi v \Big(-\frac{y}{\phi\sigma(\boldsymbol{\theta}^T\boldsymbol{x})} + \frac{1-y}{1-\phi\sigma(\boldsymbol{\theta}^T\boldsymbol{x})}\Big) \cdot p_z(b(f_{\boldsymbol{\theta}}(\boldsymbol{x}))) \cdot \sigma(\boldsymbol{\theta}^T\boldsymbol{x})(1 - \sigma(\boldsymbol{\theta}^T\boldsymbol{x}))\boldsymbol{x}$ |

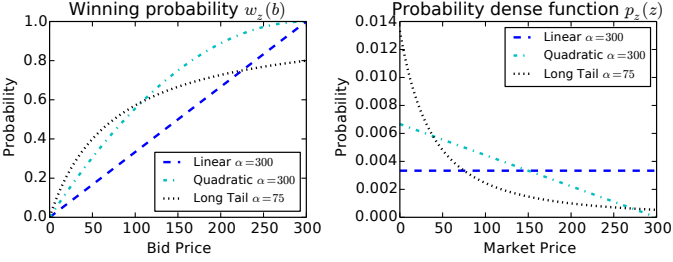

Fig. 4: An illustration of landscapes with different $\alpha$ values, the forms of $p_z(z)$ reflecting the analysis results in Sec. 7.6

## 5 COST: MARKET COMPETITION MODELING

From the above derivation, the market price distribution $p_z(z)$ has great contributions to the utility estimation component. Moreover, as is mentioned in Sec. 4.3, the landscape function $p_z(z)$ has various realizations. In this section, we propose a machine learning methodology to model the market competition, which is to learn the market price distribution function $p_z(z, \boldsymbol{x}; \boldsymbol{\phi})$ with the specified feature $\boldsymbol{x}$. Besides, our learning objective is the profit of the advertiser, rather than merely the likelihood between the true distribution and the learned model, though our model has similar prediction accuracy of distribution modeling as shown in the experiments in Appendix E.

We take the objective function of expected utility as Eq. (59) and the gradient of $R$ w.r.t. $\boldsymbol{\phi}$ can be derived as

$$\frac{\partial R}{\partial \boldsymbol{\phi}} = \frac{\partial}{\partial \boldsymbol{\phi}}\Big[vy\int_0^b p_z(z)dz - \int_0^b z p_z(z) dz\Big]. \quad (21)$$

Next, we will discuss several formulations of the market price distribution $p_z(z, \boldsymbol{x}; \boldsymbol{\phi})$.

The market price $z$ is a positive variable and its p.d.f. is $p_z(z), z \in [0, +\infty)$. Here we model $p_z(z, \boldsymbol{x}; \boldsymbol{\phi})$ and naturally get the winning probability $w_z$ which describes the winning probability of proposing $b$ under the market price assumption of $p_z(z)$ as in Eq. (1). Our another goal is to find the proper formulation for winning probability function $w$, and the formulated function should satisfy two properties:

$$\begin{aligned} b \to 0^+, \ w \to 0 \ ; \\ b \to +\infty, \ w \to 1 \ . \end{aligned} \quad (22)$$

Note that the p.d.f. of the market price has a long-tail property, which is shown in our discussion of Sec. 7.6, so we propose three formulations for the bid landscape function for market price modeling.

### 5.1 Linear Form

First, we base on the assumption of the uniform market price distribution ranging in $[0, \alpha(\boldsymbol{x}; \boldsymbol{\phi})]$ where $\alpha$ is the maximal value of the market price, and we take $\alpha(\boldsymbol{x}; \boldsymbol{\phi}) = e^{\boldsymbol{\phi}^T\boldsymbol{x}}$ for each bid request feature $\boldsymbol{x}$, where the exponential function is used to make sure the upperbound is positive. Thus the corresponding winning probability for the proposed bid price $b$ is

$$\begin{aligned} w(b, \boldsymbol{x}; \boldsymbol{\phi}) &= \frac{b}{\alpha(\boldsymbol{x}; \boldsymbol{\phi})} = \frac{b}{e^{\boldsymbol{\phi}^T\boldsymbol{x}}}, b \in [0, \alpha(\boldsymbol{x}; \boldsymbol{\phi})], \\ p(z, \boldsymbol{x}; \boldsymbol{\phi}) &= \frac{\partial w(z, \boldsymbol{x}; \boldsymbol{\phi})}{\partial z} = \frac{1}{e^{\boldsymbol{\phi}^T\boldsymbol{x}}} \ . \end{aligned} \quad (23)$$

Thus the gradient $\frac{\partial R}{\partial \boldsymbol{\phi}}$ derived from Eq. (21) is that

$$\begin{aligned} \frac{\partial R}{\partial \boldsymbol{\phi}} &= \frac{\partial}{\partial \boldsymbol{\phi}}\Big[vy \int_0^b p_z(z)dz - \int_0^b z p_z(z)dz\Big] \\ &= \Big(\frac{b^2}{2} - vyb\Big)\frac{\boldsymbol{x}}{e^{\boldsymbol{\phi}^T\boldsymbol{x}}} \ . \end{aligned} \quad (24)$$

As is shown in Figure 4, the winning probability function $w_z(b)$ is proportional to the bid price and the maximal market price is $\alpha$, whose value varies for different $\boldsymbol{x}$ and is learned by our parametric model.

### 5.2 Quadratic Form

We keep modeling the maximal market price as $\alpha(\boldsymbol{x}; \boldsymbol{\phi})$, that is $z \in [0, \alpha(\boldsymbol{x}; \boldsymbol{\phi})]$, and take a quadratic formulation for winning probability function $w_z(b)$. Then

$$\begin{aligned} w(b, \boldsymbol{x}; \boldsymbol{\phi}) &= \frac{b}{\alpha(\boldsymbol{x}; \boldsymbol{\phi})}\Big(2 - \frac{b}{\alpha(\boldsymbol{x}; \boldsymbol{\phi})}\Big), b \in [0, \alpha(\boldsymbol{x}; \boldsymbol{\phi})] \ , \\ p(z, \boldsymbol{x}; \boldsymbol{\phi}) &= \frac{\partial w(z, \boldsymbol{x}; \boldsymbol{\phi})}{\partial z} = -\frac{2z}{\alpha(\boldsymbol{x}; \boldsymbol{\phi})^2} + \frac{2}{\alpha(\boldsymbol{x}; \boldsymbol{\phi})}, \end{aligned} \quad (25)$$

since winning probability has the properties that $w(b=0) = 0, w(b=\alpha) = 1$, and is monotonously increasing over $[0, \alpha]$, as in illustrated in Figure 4.

Thus the gradient $\frac{\partial R}{\partial \boldsymbol{\phi}}$ derived from Eq. (21) is

$$\begin{aligned}\frac{\partial R}{\partial \boldsymbol{\phi}} &= \frac{\partial}{\partial \boldsymbol{\phi}}\Big[vy\int_0^b p_z(z)dz - \int_0^b zp_z(z)dz\Big]\\ &= \Big[vy\Big(\frac{2b^2}{\alpha(\boldsymbol{x};\boldsymbol{\phi})^3} - \frac{2b}{\alpha(\boldsymbol{x};\boldsymbol{\phi})^2}\Big) - \Big(\frac{4b^3}{3\alpha(\boldsymbol{x};\boldsymbol{\phi})^3} - \frac{b^2}{\alpha(\boldsymbol{x};\boldsymbol{\phi})^2}\Big)\Big]\\ &\quad \cdot \frac{\partial \alpha(\boldsymbol{x};\boldsymbol{\phi})}{\partial \boldsymbol{\phi}}\end{aligned} \quad (26)$$

And $\alpha(\boldsymbol{x}; \boldsymbol{\phi})$ is a functional mapping from $\boldsymbol{x}$ to the maximal market price for each bid request. We also take a regression model to fit the mapping function as that in Sec. 5.1 whose parameter is $\boldsymbol{\phi}$.

### 5.3 Long Tail Form

Some simple winning probability function [48] can be derived from Eq. (21) as

$$w(b, \boldsymbol{x}; \boldsymbol{\phi}) = \frac{b}{b + \alpha(\boldsymbol{x}; \boldsymbol{\phi})} , \quad (27)$$

we also take $\alpha(\boldsymbol{x}; \boldsymbol{\phi}) = e^{\boldsymbol{\phi}^T \boldsymbol{x}}$ as to learn the optimal parameter $\alpha$ value for each feature $\boldsymbol{x}$.

The market price distribution $p_z(z, \boldsymbol{x}; \boldsymbol{\phi})$ is

$$p_z(z, \boldsymbol{x}; \boldsymbol{\phi}) = \frac{\partial w(z, \boldsymbol{x}; \boldsymbol{\phi})}{\partial z} = \frac{e^{\boldsymbol{\phi}^T \boldsymbol{x}}}{(z + e^{\boldsymbol{\phi}^T \boldsymbol{x}})^2} . \quad (28)$$

Thus we can derive the gradient of $U$ w.r.t. $\boldsymbol{\phi}$ as

$$\begin{aligned}\frac{\partial R}{\partial \boldsymbol{\phi}} &= \frac{\partial}{\partial \boldsymbol{\phi}}\Big[vy\int_0^b p_z(z)dz - \int_0^b zp_z(z)dz\Big]\\ &= \frac{\partial}{\partial \boldsymbol{\phi}}\frac{vyb}{b + e^{\boldsymbol{\phi}^T \boldsymbol{x}}} - \frac{\partial}{\partial \boldsymbol{\phi}}\int_0^b zp_z(z)dz\\ &= \boldsymbol{x}\frac{vybe^{\boldsymbol{\phi}^T \boldsymbol{x}}}{(b + e^{\boldsymbol{\phi}^T \boldsymbol{x}})^2} + \frac{\partial}{\partial \boldsymbol{\phi}}\Big[\Big(\frac{e^{\boldsymbol{\phi}^T \boldsymbol{x}}}{e^{\boldsymbol{\phi}^T \boldsymbol{x}} + b} + \ln(e^{\boldsymbol{\phi}^T \boldsymbol{x}} + b)\\ &\quad - (\boldsymbol{\phi}^T \boldsymbol{x} + 1)\Big)e^{\boldsymbol{\phi}^T \boldsymbol{x}}\Big].\end{aligned} \quad (29)$$

As is also shown in Figure 4, the market price probability $p_z(z)$ is monotonously decreasing while $z$ is rising and has long-tail property for large market prices. Thus the winning probability function $w_z(b)$ approaches to 1 when the bid price $b$ gets larger.

### 5.4 Double Optimization for Campaign Performance

With the derived the optimal bid landscape forecasting model for campaign profits, we can naturally propose a double optimization algorithm. In Algorithm 1, both the CTR estimation model and the bid landscape forecasting model are learned, whose parameters are $\boldsymbol{\theta}$ and $\boldsymbol{\phi}$, respectively. Moreover, the bid landscape model derives the corresponding winning probability function $w_{\boldsymbol{\phi}}(b)$ w.r.t. the proposed bid price $b$.

Generally speaking, this double optimization aims to optimize the campaign profit in the view of the two prediction model. Note that, in this procedure, the contribution made by the bid landscape forecasting model $\boldsymbol{\phi}$ lies in the learning of the CTR estimation model $\boldsymbol{\theta}$ since the update of the CTR model contains the landscape forecasting results $p_z(z, \boldsymbol{x}; \boldsymbol{\phi})$ as in Eqs. (8) and (12).

---

**Algorithm 1** Double Optimization for Campaign Profits

**Input:** Training set $D$
**Output:** winning function $w_{\boldsymbol{\phi}}()$,
  Optimal utility (CTR) estimation model $f_{\boldsymbol{\theta}}()$
1: Initially set parameter $\boldsymbol{\theta}$ and $\boldsymbol{\phi}$
2: **for** num. of training rounds **do**
3:   **for** each sample $(\boldsymbol{x}, y) \in D$ **do**
4:     Calculate the gradient of $\boldsymbol{\theta}$ via Eq. (8)
5:     Calculate the gradient of $\boldsymbol{\phi}$ via Eq. (21)
6:     Update parameters $\boldsymbol{\theta}$ and $\boldsymbol{\phi}$ via gradient descent
7:   **end for**
8: **end for**

---

## 6 BID OPTIMIZATION

In this section, we will focus on bid optimization considering budget constraints. The rational is that, since the CTR estimation and winning probability have been settled down, the bidding function $b(\cdot)$ should be fine tuned for bid optimization. The click maximization framework [48] only takes the bid price as the upper bound of the cost, which is suitable in the first-price auction rather than the commonly used second-price auctions.

As is proved in Appendix B, the linear bidding function $b(u)$ w.r.t. the utility $u$ is the optimal bidding strategy under the second-price auction. Here we replace $f(\boldsymbol{x})$ with $r$ to represent the predicted user response for simplicity, and consider the linear bidding function

$$\begin{aligned}u(r) &= vr ,\\ b(r) &= \gamma u(r) = \gamma vr .\end{aligned} \quad (30)$$

Then we derive the optimal solution of parameter $\gamma$.

$$\begin{aligned}&\arg\max_{\gamma} T \int_r \int_0^{\gamma vr}(vr - z)p_z(z)dz \cdot p_r(r)dr\\ &\text{s.t. } T\int_r \int_0^{\gamma vr} zp_z(z)dz \cdot p_r(r)dr = B .\end{aligned} \quad (31)$$

The Lagrangian $\mathcal{L}(\gamma, \lambda) =$

$$T\int_r \int_0^{\gamma vr}[vr - (\lambda + 1)z]\,p_z(z)dz \cdot p_r(r)dr + \lambda B, \quad (32)$$

where $\lambda$ is the Lagrangian multiplier. Taking the derivative equal to zero, we get that

$$\frac{\partial \mathcal{L}(\gamma, \lambda)}{\partial \gamma} = 0 \Rightarrow \gamma = \frac{1}{\lambda + 1}. \quad (33)$$

To solve $\lambda$, we take the Lagrangian derivative w.r.t. to $\lambda$ ant let it be zero, which obtains the constraint equation

$$T\int_r \int_0^{\frac{vr}{1+\lambda}} zp_z(z)dz\, p_r(r)dr = B, \quad (34)$$

which normally has no analytic solution of $\lambda$ except for some trivial implementation of $p_z(z)$ and $p_r(r)$. Fortunately, the numeric solution of $\lambda$ is easy to find because the left part of the equation monotonously decreases against $\lambda$ in the bidding function.

From Eq. (34), we find that the distribution of the predicted CTR $p_r(r)$ directly influences the optimal value of $\lambda$ in the bidding function Eq. (33). It means if we update the CTR estimation model $f_{\boldsymbol{\theta}}(\boldsymbol{x})$, $p_r(r)$ will change accordingly, which in turn leads to the change of optimal $\lambda$ [48].

**Algorithm 2** Periodic Bidding Machine

**Input:** Training sets $\{D_1, D_2, \ldots, D_n\}$, total budgets $\{B_1, B_2, \ldots, B_n\}$ for each training set $D_i$
**Output:** Optimal utility (CTR) estimation model $f_{\boldsymbol{\theta}}()$, winning function $w_{\boldsymbol{\phi}}()$ and bidding strategy $b(\cdot)$
1: Initially set parameter $\boldsymbol{\theta}$ and $\boldsymbol{\phi}$ and $\gamma$
2: **for** each dataset $D_i$ in the training data **do**
3:    **for** each sample $(\boldsymbol{x}, y) \in D_i$ **do**
4:       Calculate the gradient via Eqs. (8) and (21)
5:       Optimize the corresponding parameters as gradient descent algorithm
6:    **end for**
7:    Update bidding function $b(\cdot)$ via solving Eq. (34)
8: **end for**

## 6.1 Bidding Machine Algorithm

Having specified the learning algorithms for utility optimization and market modeling, and gone through bid optimization, we deliver our bidding machine algorithm as in Algorithm 2, which is also illustrated in Figure 1. After received a bid request, our bidding machine will put it through the whole predicting process, i.e., (i) estimate utility and (ii) predict the winning function, and then (iii) propose the corresponding bid price according to the optimal bidding strategy. Our predicted results will be aligned with the ground truth, that is, if winning, the true user response to supervise the utility estimation module, and the real market price to correct the winning function learning. After that, we will fine-tune our bidding strategy to gain the optimal bidding function to maximize the advertiser's total profit. The offline training involves a joint optimization of the three components by alternatively optimizing one and fixing the other two.

We can find that the bidding machine algorithm will digest the historical bidding information and update the parameters as a whole. We have also derived the online FTRL learning paradigm [25] for bidding machine framework, which is included in Appendix A. For comparison with our previous work [33], we implement SGD learning paradigm in our offline experiments.

## 7 EXPERIMENTS

In this section, we first present the datasets and the experiment settings with evaluation metrics. Second, we will present the user response prediction model performance and bid landscape forecasting results, and discuss some reasons behind the improvement of our models. Third, we will discuss the experimental results for the whole bidding machine framework and finally the online A/B test results.

## 7.1 Datasets

We use two real-world datasets: iPinYou and YOYI, and provide repeatable offline empirical studies[2].

**iPinYou** is a leading DSP company in China. The iPinYou dataset[3] was released to promote the research on real-time bidding. The entire dataset contains 65M bid records including 20M impressions, 15K clicks and 16K CNY expense on 9 different campaigns over 10 days in 2013. The auctions during the last 3 days are set as test data while the rest as training data.

**YOYI** runs a major DSP focusing on multi-device display advertising in China. YOYI dataset[4] contains 402M impressions, 500K clicks and 428K CNY expense during 8 days in Jan. 2016. The first 7 days in the time sequence are set as the training data while the last 1 day is as the test data.

For the repeatable experiments, we focus on our study on iPinYou dataset. Our algorithms are further evaluated over the YOYI dataset for multi-device display advertising.

In real-time bidding, the training data contains much fewer positive samples than negative ones. Thus similar to [14], the negative down-sampling and the corresponding calibration methods are adopted in the experiment. The online A/B test is conducted on an operational real-time bidding platform run by YOYI.

## 7.2 Experiment Setup

**Experiment Flow.** We take the original impression history log as full volume bid request data. The data contain a list of bid record triples with user response (click) label, the corresponding market price and the request features. We follow the previous work [48] for feature engineering and the whole experiment flow, which is as follows: the bid requests are received along with the time sequence, which is the same as the procedure that history log was generated. When received one request, our bid engine will decide the bid price to participate the real-time bidding auction. It wins if its bid price is higher than the market price, otherwise loses. Note that, the overall objective is to gain as much profit as possible, and the contribution of the market modeling to the final goal is to help the learning of CTR model which sequentially affects the final bid price. Thus our user response prediction and market modeling are both derived to a CTR estimation model in the experimental measurement as results. Thus, on one hand, we deploy different CTR estimation models to predict the user response probability, which then can be compared against each other. On the other hand, we also present the experimental results of our bid landscape forecasting model with other prediction models in Appendix E. After bidding, the labeled clicks of the winning impressions will act as user feedback information. It is worth mentioning that this evaluation methodology works well for evaluating user response prediction and bid optimization [3], [48] and has been adopted in display advertising industry [21].

It is obvious that if our bid engine bids very high price each time, the cost and profit will stay the same as the original test log. Thus the budget constraints play a key role in evaluation [48]. For CTR estimation and bid landscape forecasting, we only report for the test results without budget constraints since we care more about the prediction performance. For the bidding strategy optimization, we follow [47], [48] to run the evaluation test using 1/64, 1/32, 1/16, 1/8, 1/4, 1/2 of the original total cost respectively in the test log as the budget constraints.

---

2. Repeatable experiment code: https://goo.gl/uCmdLR.
3. iPinYou Dataset link: http://goo.gl/9r8DtM.
4. YOYI Dataset link: http://goo.gl/xaao4q.

## 7.3 Evaluation Measures

Since our objective is to improve the profit of a performance campaign and cut down the unnecessary cost in bidding, in our evaluation we measure **profit** and **ROI** w.r.t the corresponding cost in bidding phase. When the bid engine wins the auction, the corresponding market price will be added into the total *cost*. While the user response (click) is positive, we will take the campaign click value (preset by the advertiser) of this action as *return*. In our settings, this click value is set equal to eCPC in the campaign's history data log. While in the real-world scenario, the campaign value is set by the advertiser. The **profit** is regarded as the total gross profit ($\sum \text{return} - \sum \text{cost}$) for the whole test data auctions. **ROI**=profit/cost is another important measurement reflecting the cost-effectiveness of a bidding strategy. It can be regarded as a relatively orthogonal metric to auction volume and bid cost.

We also take ad related metrics such as **eCPC**, cost per thousand impressions (**CPM**), **CTR**, and the **winning rate** to compare the bidding performance of the different prediction models.

To measure the binary classification performance for CTR estimation, we adopt commonly used **AUC** (area under ROC curve) [5] and **RMSE** (root mean squared error) to measure the accuracy of a regression model. Moreover, for bid landscape forecasting, we report the **ANLP** (averaged negative log probability) [40] of the forecasting results over the test dataset. The detailed experimental results of AUC, RMSE and ANLP can be found in Appendix D and E. We also present the significance test results in Appendix F.

## 7.4 Compared Settings

**Test Settings without Budget Constraint.** For the first part of our experiment, the unlimited budget is tested. All the CTR models are embedded with the same truthful bidding function. We compare 4 models in this part:

- **CE** - The logistic regression model [14], [25] is widely used in many DSP platforms to make predictions of user feedback. This model takes cross entropy as its optimization objective and has the gradient as Eq. (17).
- **SE** - This logistic regression model takes the squared error loss as the objective function, which takes the gradient update as Eq. (16).
- **EU** - Our proposed expected utility model for CTR estimation, which takes the gradient update as Eq. (14).
- **RR** - Our proposed risk-return model for CTR estimation, which takes the gradient update as Eq. (15).
- **BM(MKT)** - Our proposed CTR estimation model with market (MKT) modeling, which is described in the binary optimization method of Algorithm 1.

Note that EU and RR models consider a statistical bid landscape function $p_z(z)$, while the last BM(MKT) model utilizes a parametric market competition modeling $p_z(z, \boldsymbol{x}; \boldsymbol{\phi})$.

**Test Settings with Budget Constraint.** In the second scenario, we evaluate different bidding strategies under budget constraints. Here we test 4 solutions:

[5]. It has been shown that AUC is equal to the probability that a regressor correctly ranks a randomly chosen positive example higher than a randomly chosen negative one.

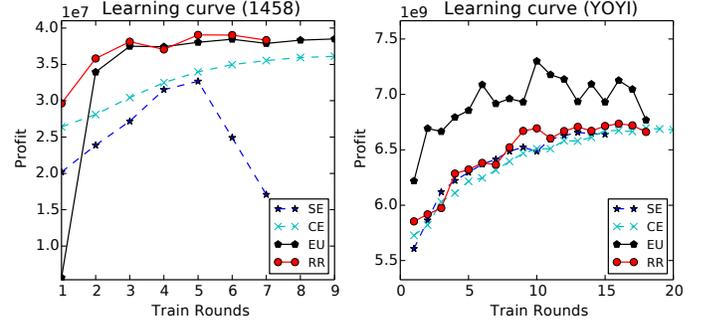

Fig. 5: Training on iPinYou (left) and YOYI (right).

TABLE 3: Direct campaign profit over baselines.

|  | Profit ($\times 10^7$) | | ROI | |
| --- | --- | --- | --- | --- |
| iPinYou | SE | CE | SE | CE |
| 1458 | 3.2 | 3.6 | 4.2 | 6.6 |
| 2259 | -0.32 | 0.40 | -0.080 | 0.18 |
| 2261 | 0.29 | 0.63 | 0.26 | 0.40 |
| 2821 | 0.11 | 0.08 | 0.21 | 0.023 |
| 2997 | 0.11 | 0.14 | 0.42 | 0.71 |
| 3358 | 1.76 | 2.4 | 5.4 | 5.2 |
| 3386 | 0.51 | 1.6 | 0.16 | 1.2 |
| 3427 | 0.33 | 2.9 | 0.11 | 3.4 |
| 3476 | 0.65 | 3.1 | 0.36 | 3.5 |
| Average | 0.74 | **1.7** | 1.2 | **2.3** |
| YOYI | 665.6 | **669.5** | 1.8 | **1.9** |

- **CELIN** - As in [30], [48], the bid value is linearly proportional to the predicted CTR. We implement a logistic regression model with a linear bidding strategy, which is widely used in lots of production environment.
- **ORTB** - Optimal Real-time Bidding strategy [48] which applies functional optimization for bidding function.
- **PRUD** - A Prudent bidder [22] which combines CTR estimation and winning price prediction together to efficiently bid in real time.
- **BM(FULL)** - The full bidding machine framework with joint optimization as described in Algorithm 2.

## 7.5 Campaign Profit Optimization

As we have found in Appendix D that our models have at least comparable performance for predicting CTR, we are now ready to examine the performance of profit optimization for each campaign in an unlimited budget setting (we will present the results under limited budget in Section 7.7.2). Figure 5 plots the obtained profit against the training rounds for the 4 models in both the iPinYou and YOYI datasets. The model learns on the whole training set in each round. While the figures show the convergence of each estimation model, SE does not well generalize its CTR prediction to the profit optimization in iPinYou dataset. Compared to RR, EU's prediction focuses on medium-valued CTR cases, which is indeed the range with high volume of clicks in YOYI's market data, while RR focuses more on higher-valued cases. This results shows EU better in winning more quality cases than RR.

We further examine the two baselines, SE and CE, with more details in Table 3. Both models achieve positive ROIs in almost all campaigns. And, in most campaigns CE outperforms SE in terms of the profit and ROI. This is consistent

TABLE 4: Campaign profit improvement over baseline CE.

|  | Profit gain | | ROI gain | |
|---|---|---|---|---|
| iPinYou | EU | RR | EU | RR |
| 1458 | 7.10% | 9.00% | 233% | 267% |
| 2259 | 81.6% | 99.3% | 233% | 472% |
| 2261 | 26.3% | 31.1% | 44.4% | 91.2% |
| 2821 | 573% | 615% | 1334% | 943% |
| 2997 | 5.00% | 0.700% | -3.60% | -11.4% |
| 3358 | 1.70% | 6.70% | 77.1% | 77.7% |
| 3386 | -1.20% | 2.50% | 20.6% | 58.3% |
| 3427 | 5.50% | 8.70% | 52.0% | 175% |
| 3476 | 4.20% | 8.60% | 16.0% | 91.1% |
| YOYI | 9.04% | 0.600% | 14.8% | 2.11% |
| Average | +71.2% | +78.2% | +202% | +217% |

TABLE 5: Overall statistics in offline evaluation.

|  | CTR ($\times 10^{-4}$) | | | | eCPC | | | |
|---|---|---|---|---|---|---|---|---|
| iPinYou | SE | CE | EU | RR | SE | CE | EU | RR |
| 1458 | 34 | 33 | 59 | 190 | 17 | 11 | 4.3 | 3.4 |
| 2259 | 3.3 | 3.6 | 3.7 | 5.8 | 303 | 235 | 172 | 136 |
| 2261 | 2.4 | 2.7 | 3.0 | 2.8 | 234 | 212 | 188 | 168 |
| 2821 | 5.5 | 5.9 | 4.8 | 7.0 | 116 | 137 | 105 | 112 |
| 2997 | 31 | 25 | 26 | 27 | 9.8 | 8.2 | 8.3 | 8.6 |
| 3358 | 51 | 41 | 69 | 61 | 18 | 19 | 12 | 12 |
| 3386 | 7.8 | 11 | 13 | 15 | 90 | 48 | 43 | 36 |
| 3427 | 7.2 | 25 | 29 | 72.8 | 98 | 25 | 17.3 | 10 |
| 3476 | 6.4 | 16 | 17 | 33.1 | 111 | 34 | 30 | 20 |
| Average | 16 | 18 | 25 | **46** | 110 | 81 | 64 | **57** |
| YOYI | 16 | 18 | **26** | 24 | 12.9 | 12.4 | **11.3** | 12 |

|  | CPM | | | | Win Rate | | | |
|---|---|---|---|---|---|---|---|---|
| iPinYou | SE | CE | EU | RR | SE | CE | EU | RR |
| 1458 | 57 | 37 | 25 | 65 | 0.22 | 0.24 | 0.13 | .041 |
| 2259 | 100 | 84 | 64 | 78 | 0.89 | 0.63 | 0.44 | 0.24 |
| 2261 | 57 | 56 | 56 | 46 | 0.55 | 0.81 | 0.71 | 0.67 |
| 2821 | 63 | 80 | 50 | 78 | 0.12 | 0.63 | 0.48 | 0.45 |
| 2997 | 30 | 20 | 21 | 22 | 0.55 | 0.63 | 0.65 | 0.63 |
| 3358 | 92 | 77 | 80 | 70 | 0.11 | 0.20 | 0.11 | 0.13 |
| 3386 | 71 | 54 | 55 | 55 | 0.82 | 0.45 | 0.36 | 0.29 |
| 3427 | 70 | 60 | 49 | 75 | 0.75 | 0.26 | 0.22 | .082 |
| 3476 | 71 | 55 | 50 | 65 | 0.49 | 0.31 | 0.31 | 0.15 |
| Average | 68 | 58 | **50** | 62 | **0.50** | 0.46 | 0.38 | 0.30 |
| YOYI | **20** | 23 | 29 | 30 | **0.36** | 0.30 | 0.22 | 0.22 |

with our finding in Appendix D that CE outperforms SE for CTR prediction accuracy.

We next pick up the best CE model and use it as the baseline to compare the profit gain and ROI gain with our proposed EU and RR models, as shown in Table 4. We can observe that (i) Both EU and RR consistently achieve higher profit than the CE baseline. Only in iPinYou campaign 3386, EU gains less profit. In average, our proposed models improve the profit about 71.2% for EU and 78.2% for RR, respectively. (ii) For the ROI metric, EU and RR get even higher overall improvements against the baseline CE. The average ROI gains are 202% for EU and 217% for RR, respectively. Those results suggest that our proposed models are much more cost-effective. (iii) RR is the best and in average it gains 7.0% and 15% than EU in profit and ROI, respectively.

Finally, Table 5 provides other statistics to summarize the overall campaign performance for the 4 CTR estimation models. CTR and eCPC reflect the quality and cost-effectiveness of the winning impressions. Our models, both EU and RR, outperform the baselines in terms of CTR and eCPC with comparable CPM and a relatively low winning rate. This indicates EU and RR successfully allocate the budget to high quality cost-effective ad inventories and avoid on the low quality ones.

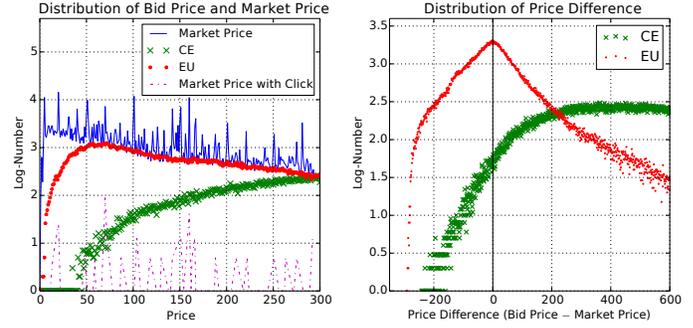

Fig. 6: Analysis of bid price and market price distribution (iPinYou campaign 2259) reflecting the formulation in Sec. 5.

TABLE 6: High bid price ($> 300$) case statistics

| Model | Auctions | Budget | Largest Bid Price |
|---|---|---|---|
| Baseline CE | 92.5% | 14.0% | 37,795 |
| Proposed Model EU | 10.3% | 1.49% | 13,901 |

### 7.6 Bidding Data Analysis

In this section, we further analyze the bidding data to gain more insights into why our models outperform the baselines. As we discussed in our formulation, a key advantage of our models is the introduction of the market price distribution and the utility of the bid to the learning of CTR model parameters. To understand the impact, in the left subfigure of Figure 6, we plot the distribution of bid values for our EU (similar to RR) model and the baseline CE model and compare them with the market price distribution and also the market price of the impressions that received clicks. We cut off the figure for price $> 300$ since the market price never goes beyond 300 in the dataset.

Firstly, we see that the bid prices generated from CE deviate far from the market prices; a large portion of the bids from CE are very high, whereas the distribution (in log scale) of the market prices gently descends from 50 to 300, with its peak in the region between 0 and 30.

By contrast, our model EU nicely reduces the *difference* between the distributions of bid price and the market price by focusing the training on the cases that the bid is close to the market price (see the discussion in Section 4.3).

Moreover, considering the market price distribution of the impressions with clicks, we find that the bid distribution of EU fits it much better than that of CE, which means the bids from EU are more unlikely to miss high quality ad impressions than those from CE.

The right subfigure in Figure 6 further shows the distributions of the price *difference* between the bids (from EU and CE respectively) and the true market prices. We find that CE has a rather biased bidding strategy — a large portion of the bids are much higher than the corresponding market prices. For EU, on the contrary, the major proportion of the bids are in the "sensitive zone" where bid price is close to the market price. The peak is located right at zero, which indicates that EU effectively leverages the market price distribution and performs sensibly.

It is particularly important to control the over spending as some of RTB auctions are in fact the first price auction or with soft floor prices [44]. Table 6 gives the statistics related to high bid price cases, where the bid value exceeds the highest market price 300 in our dataset. Specifically, for the baseline CE there are 14.0% winning auctions with bid value exceeding 300. By contrast, our model EU substantially reduces the number of high bids and controls the high price auctions fewer than 1.5% in the whole bidding process.

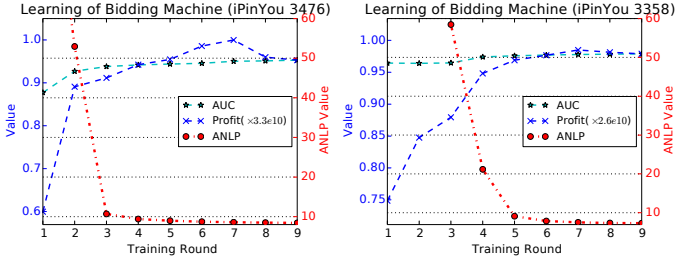

Fig. 7: Bidding Machine Learning Performance.

## 7.7 Bidding Machines

### 7.7.1 Profit Optimization with Market Modeling

In this section, we present the performance of our CTR estimation model with *learning* of the bid landscape forecasting. In Algorithm 1, the learning of bid landscape forecasting $p_z(z, \boldsymbol{x}; \phi)$ and the corresponding winning probability function $w(b, \boldsymbol{x}; \phi)$ contribute to the CTR estimation model $f_\theta(\boldsymbol{x})$. Thus, for evaluation, we implement the learned CTR model which optimizes the campaign profits with the benefits from the bid landscape forecasting model. In this experiment, we try to demonstrate that the binary optimization (with both CTR optimization and market price learning) for campaign profits is stronger than the single optimizer (with only CTR optimization).

Table 7 summarizes the results over four key metrics of the binary optimization framework which takes parametric market modeling into consideration. In the table, we can easily find that the binary optimization i.e., BM(MKT) performs the best in almost all the measurements. The improvement is reasonable since the both the CTR estimation part and the bid landscape modeling part are learned by maximizing the profits. For AUC performance, we can see that the learning of bid landscape model contributes to CTR estimation model in classification accuracy, which reflects the effectiveness of learning paradigm for the market sensitivity $p_z(z)$ in Eq. (8).

Recall that EU and RR both significantly improved the profit and ROI against normal CTR estimation model CE, as described above. The binary optimization model with market modeling makes steps further and achieve higher profits and ROI than the solely CTR learning.

Moreover, there are three main metrics for the three subproblems that are AUC for CTR estimation, ANLP for bid landscape forecasting and profit for the bidding strategy. We plot the learning curves in terms of these metrics for bidding machine as in Figure 7. We can observe that for the campaign profit, the model converges around 8 rounds, and for both AUC and ANLP, the model converges after 6 rounds, which shows the good convergence property of bidding machine.

### 7.7.2 Experiments with Budgets

As formulated in Eq. (6), the bidding machine algorithm is capable of jointly optimizing all the CTR estimation, bid landscape forecasting and bidding function by alternatively fixing two of them and optimizing the third module. In this section, we evaluate our joint optimization models under budget constraints. We mainly compare four models: CELIN, ORTB, PRUD and our integrated algorithm BM(FULL) as discussed in Section 7.4. And we set the test budget as 1/2, 1/4, 1/8, 1/16, 1/32 and 1/64 of the original total cost in the history log respectively.

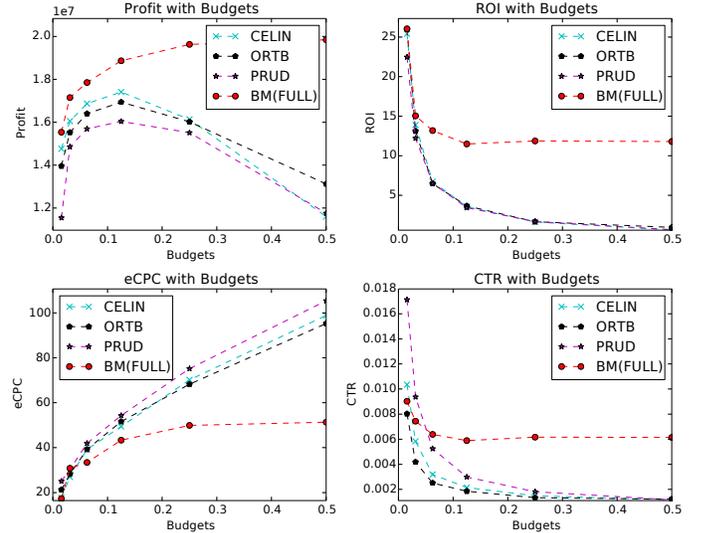

Fig. 8: Performance with budgets on iPinYou.

In Figure 8, we compare the overall performance for those three models over the tested campaigns of iPinYou dataset. The x-axis indicates the test proportion of the total cost as the budget settings. We find that in almost all settings and metrics, our proposed bidding machine algorithm BM(FULL) outperforms the other strong baselines, including state-of-the-art ORTB and PRUD.

Specifically, we find that (i) BM(FULL) achieves higher profits than other models since the learning objective of bidding machine is directly the profit. Moreover, when the budget constraint looses, the baseline models get high profits in medium degree of the constraint but drop quickly as the budget gets larger, which means that these models do not care much about the bidding efficiency and waste the budget on the high market price requests with low returns. (ii) The results of ROI and eCPC performance reflect the cost-effectiveness. We can see the overall ROI decreases to almost zero for the baselines while BM(FULL) stays effective under all settings, which means our model can dynamically drop the bid traffic with low benefits and save the budget. (iii) As for CTR, we find that PRUD achieves the highest CTR with tight budgets and our model gets better when the budget constraint is loose. The PRUD model is to maximize the total clicks and achieves state-of-the-art performance. However, the profit is not the optimization goal so that PRUD has higher CTR but lower ROI than our model.

In Table 8, we list the details of the achieved profits by all the compared models under three different budget constraints on iPinYou campaigns. The results clearly present the trends as observed in Figure 8.

TABLE 7: Campaign profit for Single CTR estimation and Binary Optimization with market modeling.

|  |  | 1458 | 2259 | 2261 | 2821 | 2997 | 3358 | 3386 | 3427 | 3476 | Average |
|---|---|---|---|---|---|---|---|---|---|---|---|
| AUC | EU | .987 | .674 | .622 | .608 | .606 | .970 | .761 | .976 | .954 | .795 |
|  | RR | .977 | .691 | .619 | .639 | .608 | .980 | .778 | .960 | .950 | .800 |
|  | BM(MKT) | .981 | .678 | .647 | .620 | .603 | .980 | .788 | .973 | .955 | **.803** |
| Profits ($\times 10^7$) | EU | 3.91 | .732 | .797 | .539 | .147 | 2.42 | 1.58 | 3.05 | 3.25 | 1.82 |
|  | RR | 3.98 | .803 | .827 | .572 | .141 | 2.54 | 1.64 | 3.14 | 3.39 | 1.89 |
|  | BM(MKT) | 4.02 | .766 | .863 | .669 | .148 | 2.57 | 1.73 | 3.18 | 3.31 | **1.91** |
| ROI | EU | 19.2 | .607 | .582 | .333 | .679 | 9.26 | 1.46 | 5.30 | 4.02 | 4.60 |
|  | RR | 24.3 | 1.03 | .771 | .247 | .624 | 9.29 | 1.90 | 9.57 | 6.63 | 6.04 |
|  | BM(MKT) | 31.7 | .829 | .692 | .476 | .733 | 8.83 | 1.08 | 9.70 | 5.40 | **6.61** |
| eCPC | EU | 4.27 | 172 | 187 | 104 | 8.33 | 11.4 | 42.5 | 17.3 | 30.0 | 64.3 |
|  | RR | 3.39 | 136 | 167 | 112 | 8.61 | 11.4 | 36.1 | 10.3 | 19.7 | **56.1** |
|  | BM(MKT) | 2.62 | 151 | 175 | 94.7 | 8.07 | 11.9 | 50.2 | 10.1 | 23.5 | 58.7 |

TABLE 8: Achieved direct profit ($\times 10^6$) with budgets.

| iPinYou | CELIN | | | ORTB | | | PRUD | | | BM(FULL) | | |
|---|---|---|---|---|---|---|---|---|---|---|---|---|
|  | 1/64 | 1/8 | 1 | 1/64 | 1/8 | 1 | 1/64 | 1/8 | 1 | 1/64 | 1/8 | 1 |
| 1458 | 40.1 | 37.1 | 22.5 | 39.9 | 37.2 | 32.8 | 35.3 | 32.3 | 3.55 | 40.7 | 40.5 | 40.5 |
| 2259 | 2.47 | 5.67 | -5.69 | 1.81 | 3.72 | -2.55 | 4.00 | 5.20 | .838 | 3.91 | 6.47 | 8.62 |
| 2261 | 1.93 | 4.48 | .151 | 1.63 | 4.99 | 3.89 | .872 | 3.47 | 2.46 | 1.97 | 5.31 | 9.13 |
| 2821 | 3.97 | 6.03 | -13.8 | 3.69 | 5.47 | -4.13 | 1.42 | 4.57 | 1.70 | 4.10 | 7.28 | 7.60 |
| 2997 | .518 | 1.25 | -1.03 | .530 | 1.36 | .151 | .111 | .665 | .675 | .670 | 1.39 | 1.49 |
| 3358 | 24.3 | 24.3 | 11.1 | 24.3 | 23.6 | 18.0 | 24.0 | 21.4 | 1.24 | 25.2 | 26.3 | 26.3 |
| 3386 | 7.83 | 13.3 | 2.43 | 6.67 | 12.3 | 10.1 | 7.10 | 11.5 | 3.71 | 8.30 | 16.4 | 18.4 |
| 3427 | 29.8 | 30.9 | 10.1 | 29.6 | 30.5 | 20.6 | 30.9 | 32.3 | 3.35 | 30.7 | 32.0 | 32.4 |
| 3476 | 21.7 | 33.3 | 15.7 | 17.3 | 33.1 | 22.8 | 19.8 | 29.6 | 2.82 | 23.3 | 33.6 | 34.8 |
| Average | 14.7 | 17.37 | 4.62 | 13.9 | 16.9 | 11.3 | 13.7 | 15.6 | 2.26 | **15.4** | **18.8** | **19.8** |

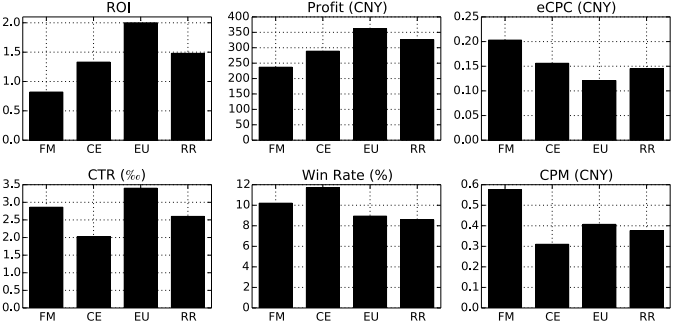

Fig. 9: Online testing results on YOYI (Phase I in 2016).

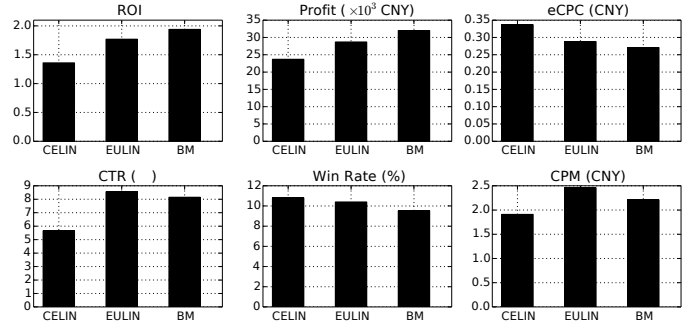

Fig. 10: Online results on YOYI MOBILE (Phase II in 2017).

## 7.8 Online A/B Testing

Our bidding machine models are deployed and tested in a live, commercial environment provided by YOYI PLUS (Programmatic Links Us) platform, which is a main DSP in China. The online experiment consists of two phases. The first phase is to test our CTR estimation models with desktop traffic, while the second phase is to compare different bidding strategies over campaigns that only target mobile inventories. The received bid requests are randomly selected to send to each model at each time according to the user cookie ID, while the chance controlled by the DSP platform for each model is set equal among all the compared models. We set the same budget constraint for all deployed models and the unit of money is CNY.

**Phase I.** We test over 10 campaigns during 25-26 January, 2016. There are 4 deployed models: EU, RR, CE and FM, where the first three have been discussed in Section 7.4 and FM is a factorization machine model [29] with non-hierarchical feature engineering. The comparison of bidding machine with our previously proposed models will be presented in Phase II. To show the comparable performance of user response prediction, we set the same linear bidding function for all prediction models including baselines. The only difference is the embedded prediction model. The whole tested bid flow contains over 89M auctions including 3.3M impressions, 8,440 clicks and 1,403 CNY budget cost. The overall results are presented in Figure 9.

From the comparison in Figure 9, we have the following conclusions: (i) EU and RR achieve higher profit and ROI than CE and FM. Specifically, EU has twice ROI as FM, and RR achieves more than 50% return of FM. EU gains 25.5% and 53.0% more profit than CE and FM respectively. (ii) eCPC consistently has inverse relationship with the trend of ROI. The online result also reflects this relationship: EU and RR have lower eCPC than other two baseline models. (iii) As for CTR, we find that EU achieves the highest CTR and RR also performs better than CE. Here FM has higher CTR than the CE model because it could learn feature interactions via the latent vector inner product [29]. However, FM obtains relatively less profit gain and ROI than CE, which shows that FM does not care enough about those auctions with high return value.

**Phase II.** We test over 5 campaigns on the mobile platform during 30 days in April, 2017. Here we deploy three bidding strategies, which are CELIN, EULIN and BM. The first two are the linear bidding strategy with CE and EU model embedded. The third algorithm is our bidding machine algorithm, namely BM in this section, which considers CTR optimization, bid landscape modeling and bidding strategy optimization altogether to maximize the whole profits. Note that, in our two-staged online testing phase, RR model performs inferior to EU model, which has been shown in Phase I. Thus we discard the RR testing in Phase II for business constraint. The whole tested bid flow involves 224M auctions including 23M impressions and 168K clicks and totally 50K CNY budget cost. The overall results are illustrated in Figure 10.

For the bidding efficiency, we have these conclusions from Figure 10: (i) Both three bidding algorithms achieves more than 100% profit w.r.t. the cost, especially the bidding machine algorithm gains almost double profit. The joint profit optimization has delivered best performance in the ROI and profit measurements. (ii) The EULIN model achieves the best (lowest) eCPC in Phase I online experiments, while our bidding machine beats the strong EULIN baseline and reduces the cost of each click in one more step. The reason is that bidding machine dynamically learns the market competition while EULIN remains static bid landscape forecasting. (iii) EULIN and BM maintain lower winning rates than CELIN while achieving higher CTR performance, which reflects the finding in Phase I online testing. (iv) BM and CELIN prefer lower price impressions as shown in CPM comparison. Moreover, our learning-based bid landscape modeling also aims to maximize the overall profits, which makes a step further than only CTR optimization and contributes to obtain more profits.

In sum, the online A/B testing results demonstrate the effectiveness of our proposed bidding machine model for profit optimization. As for the difference between offline and online experimental results, it is reasonable because of the offline data bias and the market dynamics [49].

## 8 CONCLUSIONS

In this paper, we proposed bidding machine, a comprehensive learning to bid framework, which aims to maximize the profit of the advertiser in real-time bidding for display advertising. In our bidding machine paradigm, the learning model consumes the recent bidding logs and jointly optimizes three components, including user response prediction, bid landscape forecasting and bidding strategy, in a unified objective function. Our mathematical derivations showed that the gradient of each component benefits from the behavior of the others. We tested our prediction model and the optimized bidding strategy with other state-of-the-art bidding algorithms under various market settings. The empirical study including both offline experiments and online A/B testing on a commercial RTB platform has verified the practical efficacy of our proposed bidding machine.

In the future work, we plan to integrate the censored learning [3], [40] paradigm for more accurate learning in bid landscape forecasting. Also, extending the BM framework to reinforcement learning settings [6] would be a promising direction. Besides, we would investigate the multi-agent bidding machine interactions and explore the potential equilibriums.

# APPENDIX A
# ONLINE LEARNING PARADIGM

To solve the learning problems in the bidding machine model, several algorithms have been proposed, such as SGD [34], FOBOS [10] and RDA [42]. In [24], the authors derived the relationship between FTRL-Proximal algorithm and other mirror descent algorithms and showed the better sparsity of it. For sequentially update the embedded model of the bidding machine, we derive FTRL-Proximal algorithm [25] to dynamically control the learning process while maintaining satisfying sparsity.

We first denote $g_t$ as the gradient of the $t^{th}$ instance. To update the CTR estimation model and winning probability model, the corresponding equation are

$$\boldsymbol{g}_t^{\boldsymbol{\theta}} = \frac{\partial U_t}{\partial \boldsymbol{\theta}}, \quad \boldsymbol{g}_t^{\boldsymbol{\phi}} = \frac{\partial U_t}{\partial \boldsymbol{\phi}}. \tag{35}$$

and $\boldsymbol{g}_{1:t} = \sum_{s=1}^{t} \boldsymbol{g}_s$.

In online gradient descent styple, the model parameters $\boldsymbol{\theta}$ and $\boldsymbol{\phi}$ will be respectively updated as

$$\begin{aligned}\boldsymbol{\theta}_{t+1} &= \boldsymbol{\theta}_t - \eta_t^{\boldsymbol{\theta}} \boldsymbol{g}_t^{\boldsymbol{\theta}}, \\ \boldsymbol{\phi}_{t+1} &= \boldsymbol{\phi}_t - \eta_t^{\boldsymbol{\phi}} \boldsymbol{g}_t^{\boldsymbol{\phi}},\end{aligned} \tag{36}$$

where $\eta_t$ is a non-increasing learning rate schedule. Instead, we use FTRL-Proximal paradigm for parameter updating:

$$\begin{aligned}\boldsymbol{\theta}_{t+1} &= \arg\min_{\boldsymbol{\theta}} (\boldsymbol{g}_{1:t}^{\boldsymbol{\theta}} \cdot \boldsymbol{\theta}_t + \frac{1}{2} \sum_{s=1}^{t} \delta_s \|\boldsymbol{\theta} - \boldsymbol{\theta}_s\|_2^2 + \lambda_1 \|\boldsymbol{\theta}\|_1), \\ \boldsymbol{\phi}_{t+1} &= \arg\min_{\boldsymbol{\phi}} (\boldsymbol{g}_{1:t}^{\boldsymbol{\phi}} \cdot \boldsymbol{\phi}_t + \frac{1}{2} \sum_{s=1}^{t} \delta_s \|\boldsymbol{\phi} - \boldsymbol{\phi}_s\|_2^2 + \lambda_1 \|\boldsymbol{\phi}\|_1),\end{aligned} \tag{37}$$

where we define $\delta_{1:t} = \frac{1}{\eta_t}$ as in terms of the learning rate schedule.

As is shown in [25], only one number per coefficient needs to be stored and we can rewrite the argmin equation as a quadratic function of the two parameter

$$\begin{aligned}&\frac{1}{\eta_t} \|\boldsymbol{\theta}\|_2^2 + (\boldsymbol{g}_{1:t}^{\boldsymbol{\theta}} - \sum_{s=1}^{t} \delta_s \boldsymbol{\theta}_s) \cdot \boldsymbol{\theta} + \lambda_1 \|\boldsymbol{\theta}\| + (\text{const}), \\ &\frac{1}{\eta_t} \|\boldsymbol{\phi}\|_2^2 + (\boldsymbol{g}_{1:t}^{\boldsymbol{\phi}} - \sum_{s=1}^{t} \delta_s \boldsymbol{\phi}_s) \cdot \boldsymbol{\phi} + \lambda_1 \|\boldsymbol{\phi}\| + (\text{const}).\end{aligned} \tag{38}$$

If we store $\boldsymbol{z}_{t-1}^{\boldsymbol{\theta}} = \boldsymbol{g}_{1:t-1} - \sum_{s=1}^{t-1} \delta_s \boldsymbol{\theta}_s$ and the same for $\boldsymbol{\phi}$, and we update at $t^{th}$ instance as

$$\begin{aligned}\boldsymbol{z}_t^{\boldsymbol{\theta}} &= \boldsymbol{z}_{t-1}^{\boldsymbol{\theta}} + \boldsymbol{g}_t^{\boldsymbol{\theta}} + (\frac{1}{\eta_t} - \frac{1}{\eta_{t-1}}) \boldsymbol{\theta}_t, \\ \boldsymbol{z}_t^{\boldsymbol{\phi}} &= \boldsymbol{z}_{t-1}^{\boldsymbol{\phi}} + \boldsymbol{g}_t^{\boldsymbol{\phi}} + (\frac{1}{\eta_t} - \frac{1}{\eta_{t-1}}) \boldsymbol{\phi}_t.\end{aligned} \tag{39}$$

So that we can update per-coordinate of the parameter as

$$\begin{aligned}\boldsymbol{\theta}_{t+1,i} &= \begin{cases} 0 & \text{if} |z_{t,i}^{\boldsymbol{\theta}}| \leq \lambda_1, \\ -\eta_t^{\boldsymbol{\theta}} (z_{t,i}^{\boldsymbol{\theta}} - sgn(z_{t,i}^{\boldsymbol{\theta}}) \lambda_1) & \text{otherwise},\end{cases} \\ \boldsymbol{\phi}_{t+1,i} &= \begin{cases} 0 & \text{if} |z_{t,i}^{\boldsymbol{\phi}}| \leq \lambda_1, \\ -\eta_t^{\boldsymbol{\phi}} (z_{t,i}^{\boldsymbol{\phi}} - sgn(z_{t,i}^{\boldsymbol{\phi}}) \lambda_1) & \text{otherwise},\end{cases}\end{aligned} \tag{40}$$

where the second subscript $i$ is the coordinate index of the parameter.

For learning rates $\eta_t$, we implement as per-coordinate rate updating as

$$\eta_{t,i} = \frac{\alpha}{\beta + \sqrt{\sum_{s=1}^{t} \boldsymbol{g}_{s,i}^2}}, \tag{41}$$

where $i$ is the index of the coordinate.

# APPENDIX B
# OPTIMALITY UNDER SECOND-PRICE AUCTION

In this section, we give the proof of the optimal bidding function under the second-price auction.

**Theorem 1.** *The optimal bidding function under the second-price auction is linear to the estimated utility.*

*Proof.* Recall that, under the second price auction, the winning probability $w(b)$ w.r.t. the bid price $b$ is integral over $[0, b]$ for the market price distribution $p_z(z)$ as

$$w(b) = \int_0^b p_z(z) dz. \quad (42)$$

And $c(b)$ is the expected cost of bidding with price $b$.

$$c(b) = \frac{\int_0^b z p_z(z) dz}{\int_0^b p_z(z) dz}. \quad (43)$$

We use $r$ to represent the predicted user response of the given bid request, while $b(r)$ is the bidding function w.r.t. $r$ and $u(r)$ is the utility function set by the advertiser.

Our optimization problem is to maximize the *profit* of the advertiser with the budget constraint $B$ under the second-price auction, which is formulated as

$$\begin{aligned}\max_{b()} \quad & T\int_r [u(r) - c(b(r))] w(b(r)) p_r(r) dr, \\ \text{s.t.} \quad & T\int_r c(b(r)) w(b(r)) p_r(r) dr = B,\end{aligned} \quad (44)$$

where $T$ is the total number of the bid requests.

The Lagrangian of the optimization problem Eq. (44) is

$$\begin{aligned}\mathcal{L}(b(r), \lambda) = & \int_r [u(r) - c(b(r))] w(b(r)) p_r(r) dr \\ & - \lambda \cdot \int_r c(b(r)) w(b(r)) p_r(r) dr + \frac{\lambda B}{T},\end{aligned} \quad (45)$$

where $\lambda$ is the Lagrangian multiplier.

**Solving $b()$.** Based on calculus of variations, the Euler-Lagrangian condition of $b(r)$ is

$$\frac{\partial \mathcal{L}(b(r), \lambda)}{\partial b(r)} = 0, \quad (46)$$

which can be derived as

$$\begin{aligned}0 = & \frac{\partial \mathcal{L}(b(r), \lambda)}{\partial b(r)} \\ \Rightarrow \quad 0 = & u(r) p_r(r) \frac{\partial w(b(r))}{\partial b(r)} - (\lambda + 1) p_r(r) \\ & \cdot \left[ \frac{\partial c(b(r))}{\partial b(r)} w(b(r)) + c(b(r)) \frac{\partial w(b(r))}{\partial b(r)} \right] \\ \Rightarrow \quad (\lambda + 1) & p_r(r) \frac{\partial c(b(r))}{\partial b(r)} w(b(r)) \\ = & [u(r) - (\lambda + 1) c(b(r))] p_r(r) \frac{\partial w(b(r))}{\partial b(r)}.\end{aligned} \quad (47)$$

Since

$$\frac{\partial c(b(r))}{\partial b(r)} = \frac{p_z(b(r)) \left[ b(r) \int_0^b p_z(z) dz - \int_0^b z p_z(z) dz \right]}{(w(b(r)))^2}, \quad (48)$$

and

$$\frac{\partial w(b(r))}{\partial b(r)} = p_z(b(r)), \quad (49)$$

taking Eq. (48) and Eq. (49) into Eq. (47), we can then derive the Euler-Lagrangian condition as

$$\begin{aligned}(\lambda + 1) b(r) &= u(r) \\ \Rightarrow \quad b(r) &= \frac{u(r)}{\lambda + 1}.\end{aligned} \quad (50)$$

∎

# APPENDIX C
# GAME THEORETIC ANALYSIS

In this section, we conduct a theoretic analysis of the optimal bidding strategy under the symmetric game of repeated auctions with budget constraints following [46], [49]. First, we will derive the optimal bidding function in the equilibrium of the second price auction. Second, based on the derived bidding function, we discuss that a *tragedy of the commons* situation exists among multiple advertisers with the same optimal bidding strategy in RTB display advertising. Note that the analysis may not be first proposed in this work and we present it here to make this paper self-contained.

## C.1 Problem Settings

At first we present some preliminaries and describe the problem settings. We add subscripts $b, z, r$ to the c.d.f. and p.d.f. functions to make differences among these variables.

**Monotonicity of the Bidding Function.** In a clean game theoretic analysis setting [26], there are $n$ ($n \geq 2$) advertisers with the same bidding strategy $b(r)$ which takes the estimated CTR $r$ and outputs the bid price $b$. It is reasonable that $b(r)$ is monotonically increasing w.r.t. CTR $r$, i.e.

$$b(r_1) > b(r_2) \Leftrightarrow r_1 > r_2. \quad (51)$$

Later we will prove this monotonicity. Each time when an impression is auctioned, for each advertiser the CTR $r$ follows the same p.d.f. $p_r(r)$ independently (i.i.d.) and the corresponding c.d.f. is $F_r(r)$ as

$$F_r(r) = \int_0^r p_r(t) dt, \quad \frac{\partial F_r(r)}{\partial r} = p_r(r). \quad (52)$$

We also define $F_b(b)$ as the c.d.f. of the bid price $b$, i.e. the probability of performing a bid less than $b$:

$$F_b(b) = \int_0^b p_b(a) da. \quad (53)$$

Note that

$$F_b(b(r)) = P(b(r) > b_2) = P(r > r_2) = F_r(r), \quad (54)$$

since $b(r)$ monotonously increases w.r.t. $r$. Thus, for the market price variable $z$, which is defined as the highest bid price across $(n-1)$ competitors, its c.d.f. $F_z(z)$ is

$$F_z(z) = F_b(z)^{n-1}, \quad (55)$$

and the corresponding p.d.f. $p_z(z)$ is

$$p_z(z) = \frac{\partial F_z(z)}{\partial z} = (n-1) F_b(z)^{n-2} p_b(z). \quad (56)$$

**The Winning Probability in a Symmetric Game.** In such a setting, the winning probability $w_r(r)$ of Advertiser 1, without loss of generality, w.r.t. the given CTR $r$ is the largest one among the $n$ advertisers that

$$w_r(r) = P(r > r_2, r > r_3, \ldots, r > r_n) = F_r(r)^{n-1}. \quad (57)$$

Note that, according to Eqs. (54) and (55), we can also derive the winning probability function $w_b(b(r))$, which is equivalent to $w(b(r))$ in the main part of our paper, w.r.t. the bid price is that

$$\begin{aligned}w_b(b(r)) &= P(b(r) > b_2, b(r) > b_3, \ldots, b(r) > b_n) \\ &= F_b(b(r))^{n-1} = F_r(r)^{n-1} \\ &= P(r > r_2, r > r_3, \ldots, r > r_n) \\ &= w_r(r).\end{aligned} \quad (58)$$

**The Expected Utility and the Expected Cost.** We follow our previously conducted results in Eq. (43) and derive the expected utility of profit $R(r, b)$ as that

$$R(r, b) = u(r) - c(b) = u(r) - \frac{\int_0^b z p_z(z) dz}{\int_0^b p_z(z) dz}. \quad (59)$$

where $b$ is the output variable of the bidding function $b(r)$.

## C.2 Optimal Bidding Function under Symmetric Game

Now that we have defined the problem settings with the utility and the cost function, we will derive the optimal bidding function under the symmetric game scenario, where each advertiser participating in the game adopts the same bidding function.

**Theorem 2.** *The optimal bidding function under a symmetric game of repeated auctions with budget constraints is linear to the estimated utility, the bid price is monotonously increasing w.r.t. the number of the participating advertiser bidders.*

*Proof.* Our optimization problem is to maximize the *profit* of each participating advertiser with the budget constraint $B$ under the second price auction, which is formulated as

$$\begin{aligned}\max_{b()} \quad & T\int_r [u(r) - c(b(\tau))] w_b(b(\tau)) p_r(r) dr, \\ \text{s.t.} \quad & T\int_r c(b(\tau)) w_b(b(\tau)) p_r(r) dr = B,\end{aligned} \quad (60)$$

here we assume that the bidding is based on a *signal* $\tau$ related with the CTR $r$.

The Lagrangian function $\mathcal{L}(\tau, \lambda)$ is

$$\begin{aligned}\mathcal{L}(\tau, \lambda) = & \frac{\lambda B}{T} + \int_r u(r) w_b(b(\tau)) p_r(r) dr \\ & - (\lambda + 1) \int_r c(b(\tau)) w_b(b(\tau)) p_r(r) dr,\end{aligned} \quad (61)$$

where $\lambda$ is the Lagrangian multiplier. Note that the utility function $u(r)$ is only influenced by the true CTR $r$ and the cost is dependent on the bid price which is based on the known CTR signal $\tau$.

Taking Eqs. (58) and (59) into consideration, the Lagrangian function can be derived as

$$\mathcal{L}(\tau,\lambda) = \frac{\lambda B}{T} + \int_r \left[ u(r) F_r(\tau)^{n-1} - (\lambda+1) \int_0^{b(\tau)} z p_z(z) dz \right] p_r(r) dr , \quad (62)$$

**Solving** $b()$. We can calculate the gradient w.r.t. $\tau$ as that

$$\frac{\partial \mathcal{L}(\tau,\lambda)}{\partial \tau} = \int_r \left[ u(r)(n-1) F_r(\tau)^{n-2} p_r(\tau) - (\lambda+1) b(\tau) p_z(b(\tau)) \frac{\partial b(\tau)}{\partial \tau} \right] p_r(r) dr . \quad (63)$$

In a symmetric equilibrium, the objective is maximized when using the true signal, i.e., at $\tau = r$ [26]. Therefore, we have

$$\left. \frac{\partial \mathcal{L}(\tau,\lambda)}{\partial \tau} \right|_{\tau=r} = 0 \Rightarrow \quad (64)$$

$$u(r)(n-1) F_r(r)^{n-2} p_r(r) = (\lambda+1) b(r) p_z(b(r)) \frac{\partial b(r)}{\partial r} .$$

As $b(r)$ is monotonously increasing w.r.t. $r$, then their p.d.f.s $p_r(r)$ and $p_b(b)$ have the following relationship

$$p_r(r) = p_b(b(r)) \frac{\partial b(r)}{\partial r} . \quad (65)$$

Taking Eqs. (56) and (65) into Eq. (64), we can have that

$$u(r)(n-1) F_r(r)^{n-2} p_r(r)$$
$$= (\lambda+1) b(r)(n-1) F_b(b(r))^{n-2} p_b(b(r)) \frac{\partial b(r)}{\partial r}$$
$$\Rightarrow u(r)(n-1) F_r(r)^{n-2} p_r(r) \quad (66)$$
$$= (\lambda+1) b(r)(n-1) F_b(b(r))^{n-2} p_r(r)$$
$$\Rightarrow b(r) = \frac{u(r)}{\lambda+1} .$$

We can easily find that the bidding function is linear w.r.t. the utility $u(r)$. Specifically, in Sec. 3.1 of our paper, we adopt a utility function as in Eq. (4) that

$$u(r) = vr , \quad (67)$$

where $v$ is the click value of the advertiser. Therefore,

$$b(r) = \frac{vr}{\lambda+1} . \quad (68)$$

Till now, we have derived the optimal bidding function under a symmetric game of repeated auctions with multiple advertisers adopting the same bidding strategy. Comparing the derived bidding function with that in the single bidder situation, the only difference is the value of $\lambda$. Next, we will illustrate that $\lambda$ has a strong relationship with the number of the participating bidders.

**Solving** $\lambda$. Take gradient of the Lagrangian function w.r.t. $\lambda$, we can get the budget constraint equation as

$$\frac{\partial \mathcal{L}(\tau,\lambda)}{\lambda} = 0$$
$$\frac{B}{T} = \int_r \int_0^{\frac{vr}{\lambda+1}} z p_z(z) dz \, p_r(r) dr$$
$$\frac{B}{T} = \int_r \int_0^{\frac{vr}{\lambda+1}} z(n-1) F_b(z)^{n-2} p_b(z) dz \, p_r(r) dr$$
$$\frac{B}{T} = \int_r \int_0^{\frac{vr}{\lambda+1}} z(n-1) F_r\left(\frac{(\lambda+1)z}{v}\right)^{n-2} p_r\left(\frac{(\lambda+1)z}{v}\right) \quad (69)$$
$$\cdot \frac{(\lambda+1)}{v} dz \, p_r(r) dr$$
$$\frac{B}{T} = \int_r \int_0^r \frac{vt}{\lambda+1}(n-1) F_r(t)^{n-2} p_r(t) dt \, p_r(r) dr$$
$$\frac{v}{\lambda+1} = \frac{B}{T \int_r \int_0^r t(n-1) F_r(t)^{n-2} p_r(t) dt \, p_r(r) dr} .$$

The third equation is derived with Eq. (56) and the forth equation considers Eq. (54). Thus we can get the optimal bidding function as

$$b(r) = \frac{vr}{\lambda+1} = \frac{Br}{T \int_r \int_0^r t(n-1) F_r(t)^{n-2} p_r(t) dt \, p_r(r) dr} . \quad (70)$$

From Eq. (70), we can easily find that the denominator is positive and its gradient w.r.t. $n \geq 2$ is negative, which means that the bid price $b(r)$ is monotonously increasing when the bidder number $n$ increases. ∎

TABLE 9: Regression performances over campaigns. AUC: the higher, the better. RMSE: the smaller, the better.

| iPinYou | AUC | | | | RMSE ($\times 10^{-2}$) | | | |
|---|---|---|---|---|---|---|---|---|
| | SE | CE | EU | RR | SE | CE | EU | RR |
| 1458 | .948 | .987 | .987 | .977 | 3.01 | 1.94 | 2.42 | 2.32 |
| 2259 | .542 | .692 | .674 | .691 | 2.01 | 1.77 | 1.76 | 1.79 |
| 2261 | .490 | .569 | .622 | .619 | 1.84 | 1.68 | 1.71 | 1.68 |
| 2821 | .511 | .620 | .608 | .639 | 2.56 | 2.43 | 2.39 | 2.46 |
| 2997 | .543 | .610 | .606 | .608 | 5.98 | 5.82 | 5.84 | 5.82 |
| 3358 | .863 | .974 | .970 | .980 | 3.07 | 2.47 | 3.32 | 2.67 |
| 3386 | .593 | .768 | .761 | .778 | 2.95 | 2.84 | 3.32 | 2.85 |
| 3427 | .634 | .976 | .976 | .960 | 2.78 | 2.20 | 2.61 | 2.34 |
| 3476 | .575 | .957 | .954 | .950 | 2.50 | 2.32 | 2.39 | 2.33 |
| Average | .633 | .794 | .795 | **.800** | 2.97 | **2.61** | 2.86 | 2.69 |
| YOYI | .882 | .891 | **.912** | .912 | 11.9 | 11.7 | 11.8 | **11.6** |

**Analytic Solution with a Special Case.** Here we propose an analytic solution for $b(r)$ with a special case of $p_r(r)$. Assume that CTR value $r$ is uniformly distributed, which means that

$$p_r(r) = 1 , F_r(r) = r . \quad (71)$$

Thus the closed form of the optimal bidding function is

$$b(r) = \frac{vr}{\lambda+1} = \frac{Br}{T \int_{r'} \int_0^{r'} t(n-1)t^{n-2}(t) dt \, dr'}$$
$$= \frac{Br}{T \int_{r'} \frac{n-1}{n} r'^n \, dr'} = \frac{Br \cdot n(n+1)}{T(n-1)} . \quad (72)$$

In this case, we find that the optimal bidding function is linear w.r.t. the average budget per auction $\frac{B}{T}$ and the number $n$ of the participating bidders in the market. When there are more than two advertisers (i.e., $n \geq 2$), the optimal bid price is monotonously increasing when $n$ increases.

### C.3 Discussion about Tragedy of the Commons

In this part, we discuss about the derived results above. We define the performance comparison scheme. First the advertiser will compare the achieved utility $u(r)$, e.g., $u(r) = \sum vy$ and $y$ is the click indicator of each ad impression. At this point, the higher utility, i.e., more gained clicks, the better; Second, if the utility values are the same, the lower cost for achieving this utility is better.

From Eqs. (70) and (72), the optimal bid price is monotonously increasing when the number of the competitor gets larger, which means that each bidder tries to maximize the objective utility with the cost lower than the budget. When the number of the participating bidders with the same optimal bidding function is $n$, each bidder will win the auction with $\frac{1}{n}$ probability. Such that each bidder will try to spend all the budget to maximize the objective utility. However, such an equilibrium is not efficient and it will result in a situation with very low social welfare since all the advertiser will exhaust all the budget while winning the same utility, i.e., $\frac{1}{n}$ impressions and clicks.

A better situation is that each advertiser spends $B/n$ budget and still gets the same utility (the same number of impressions and clicks as in the previous case). Extremely when $n \to \infty$, each advertiser bids 0 so that the winner will be selected randomly across all the bidders pays 0 for each auction. However, this situation is never realistic since each bidder will compete with each other rather than cooperation. In such an unstable case, every advertiser will propose higher bid price to win the auction to maximize the expected utility given the current market situation, i.e., $F_z(z)$. Finally, the whole system will get into the equilibrium of Eqs. (70) and (72) where every advertiser spend out all the budgets.

This is a *tragedy of the commons* [16] reflecting a *prisoner's dilemma* [32] in the RTB market competition with budget constraints. Note that such a tragedy of the commons result is not just for the proposed bidding strategies in this work, it applies to any bidding fuctions that is monotonously increasing w.r.t. the expected utility. More discussions are provided in [46].

## APPENDIX D
## ACCURACY COMPARISON OF CTR ESTIMATION

In this section, we compare the accuracy of the CTR estimation models, measured by AUC and RMSE. As our utility estimation models are designed to optimize campaign profit rather than user response prediction accuracy, the evaluation here is to see whether our proposed solutions would still be able to achieve comparable performance against the conventional estimators that directly optimize the prediction accuracy. Table 9 shows the AUC and RMSE for each model over all campaigns. First, the baseline CE achieves better performance than the baseline SE on all campaigns, confirming the previous study that cross entropy as an objective naturally works better on the binary classification problem with probabilistic predictions. Second, both our EU and RR models achieve similar or higher AUC values over the strong baseline CE model, while maintaining comparable RMSE performances. From our derivation in the main theory section, we know that a key advantage of our EU model over the baseline SE model is that it considers the market price in the gradient updating. Here, we find that our EU model not only compensates the relatively weakness of the SE model, but also gains better in some campaigns, e.g., iPinYou campaign 2261 and YOYI. Moreover, the EU model achieves similar (sometimes better) performances compared with the CE model. Finally, we also observe that our RR model performs more stably in most campaigns and achieves higher AUC than other three models in most campaigns, e.g., iPinYou campaigns 2821, 3358, 3386 and YOYI, suggesting that combining the cross entropy loss with the market price density is the best option.

TABLE 10: Experimental results of bid landscape forecasting models over iPinYou dataset. ANLP: The lower, the better.

| Model | 1458 | 2259 | 2261 | 2821 | 2997 | 3358 | 3386 | 3427 | 3476 |
|---|---|---|---|---|---|---|---|---|---|
| NM | 5.36 | 6.76 | 5.53 | 6.55 | 5.36 | 5.83 | 5.27 | 4.88 | 5.28 |
| MM | 5.78 | 7.32 | 7.02 | 7.26 | 6.70 | 7.17 | 6.14 | 6.18 | 6.02 |
| Linear | 7.54 | 7.38 | 7.28 | 7.27 | 7.28 | 7.54 | 7.50 | 7.66 | 8.01 |
| Quadratic | 6.56 | 9.18 | 9.15 | 11.4 | 6.48 | 8.50 | 6.84 | 6.95 | 6.64 |

TABLE 11: $p$-values under the AUC evaluation (MannWhitney U test, one-tailed).

| Models | EU | RR | BM(MKT) |
|---|---|---|---|
| SE | $< 10^{-6}$ | $< 10^{-6}$ | $< 10^{-6}$ |
| CE | 0.442676 | 0.192533 | 0.09581 |

# APPENDIX E
# RESULTS OF MARKET COMPETITION MODELING

In this section, we present the experimental results of our market competition model, which is learned landscape p.d.f. $p_z(z, \boldsymbol{x}; \boldsymbol{\phi})$.

The metric is the Averaged Negative Log Probability (ANLP) [40], which is to measure the averaged log-likelihood of fitting the observed market price in the test data:

$$P_{NL} = -\frac{1}{n} \sum_{i=1}^{n} \log p_z(z_i, \boldsymbol{x}_i; \boldsymbol{\phi}) , \qquad (73)$$

where $i$ is the index of the sample in the test dataset and $p_z(z_i, \boldsymbol{x}_i; \boldsymbol{\phi})$ is the corresponding probabilistic density calculated by the bid landscape model. The better fitting performance is, the lower ANLP value it achieves.

We report the ANLP performance over the market modeling model in our paper and the state-of-the-art model: the Mix Model (MM) [41] using linear regression and censored regression altogether, and the normal model (NM) as in [40] using only observed data without lost censored data. Our model includes the performance of linear form and quadratic form of market modeling function, as described in our paper.

As is illustrated in Table 10, we find that the NM model achieves the best ANLP results and our models are in the same level as the MM model. The results are reasonable since NM fits the raw market price distribution better while MM pays more attention on the lost auctions which are censored in the true market. Our model not only optimizes the direct profits but also learns the bid landscape information very well.

# APPENDIX F
# SIGNIFICANCE TEST

In this section, we present the results of the significance test in our experiments.

For the AUC metric, we conducted a MannWhitney U test [23] and list the results in Table 11. We find that our proposed models has significantly beaten the linear regression model with squared loss (SE). However, the linear regression with cross entropy loss (CE) has similar AUC performance with our models. It is reasonable since our models aim at optimizing the overall revenue of the advertiser, rather than the classification accuracy.

For the RMSE metric, we tested $p$-values [5] of the predicted CTR values across all the models, which is illustrated in Table 12. From the results we can find that the improvement of EU and RR model over linear regression models are significant, either for the BM(MKT) model against all the baselines.

We also report the results of $p$-values under ANLP metric, which is shown in Table 13. The results have shown the significant results for our bid landscape modeling against other baselines.

TABLE 12: $p$-values under the RMSE evaluation.

| Models | EU | RR | BM(MKT) |
|---|---|---|---|
| SE | $< 10^{-6}$ | $< 10^{-6}$ | $< 10^{-6}$ |
| CE | $< 10^{-6}$ | $< 10^{-6}$ | $< 10^{-6}$ |

TABLE 13: $p$-values under the ANLP Metric.

| Models | Linear | Quadratic |
|---|---|---|
| NM | $< 10^{-6}$ | $< 10^{-6}$ |
| MM | $< 10^{-6}$ | $< 10^{-6}$ |

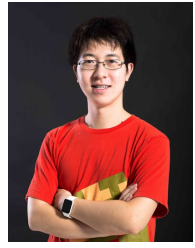

**Kan Ren** is a PhD student in Department of Computer Science in Shanghai Jiao Tong University. His research interests include data mining, machine learning, reinforcement learning and applied data science including computational advertising and sequential decision making. He has published several research papers in KDD, CIKM, ECML-PKDD, WSDM and ICDM.

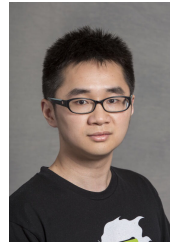

**Weinan Zhang** is an assistant professor in Department of Computer Science, Shanghai Jiao Tong University. His research interests include machine learning and data mining, particularly, deep learning and reinforcement learning techniques for real-world data mining scenarios. He has published 50 research papers on conferences & journals including KDD, SIGIR, AAAI, WWW, WSDM, ICDM, JMLR and IPM etc.

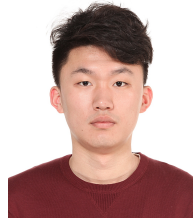

**Ke Chang** is a master student in Carnegie Mellon University, School of Computer Science, and majors in computational data science. He has earned his bachelors degree in Computer Science in Shanghai Jiao Tong University. His research interests include machine learning, data mining and the corresponding system development.

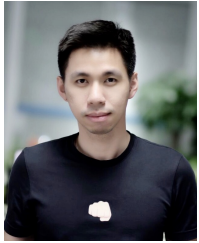

**Yifei Rong** is responsible for the algorithm R&D in Meituan-Dianping Inc. He has led the research of online advertising in a top DSP company in China. His research interests include user response learning, user profiling and bid optimization and he has also published several scientific papers about computational advertising in WSDM and CIKM.

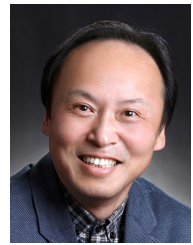

**Yong Yu** is a professor in Department of Computer Science in Shanghai Jiao Tong University. His research interests include information systems, web search, data mining and machine learning. He has published over 200 papers and served as PC member of several conferences including WWW, RecSys and a dozen of other related conferences (e.g., NIPS, ICML, SIGIR, ISWC etc.) in these fields.

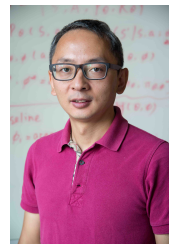

**Jun Wang** is a professor of information systems and data science in University College London. He has published over 100 research papers in leading journals and conference proceedings including TOIS, TKDE, WWW, CIKM, SIGIR, KDD, SIGMM, AAAI etc. He received the Best Doctoral Consortium award in ACM SIGIR'06, the Best Paper Prize in ECIR'09, ECIR'12, ADKDD'14 and SIGIR'17.